\documentclass{article}
\usepackage{graphicx}
\begin{document}
\begin{center}{\Large Accelerating Universe: Theory versus Experiment}
\end{center}

\begin{center}
{Moshe Carmeli $^\star$}
\end{center}

\begin{center}
{Department of Physics, Ben Gurion University, Beer Sheva 84105, 
Israel}
\end{center}

\renewcommand{\abstractname}{}
\begin{abstract}
The theory presented here, cosmological general relativity, uses a Riemannian 
four-dimensional presentation of 
gravitation in which the coordinates are those of Hubble, i.e. distances and 
velocity rather than the traditional space and time. We solve the field 
equations and show
that there are three possibilities for the Universe to expand. The theory 
describes the Universe as having a three-phase evolution with a decelerating
expansion, followed by a constant and  an accelerating expansion, and it 
predicts that the Universe is now in the latter phase. It is
shown, assuming $\Omega_m=0.245$, that the time at which the Universe goes over
from a decelerating to an accelerating expansion, i.e., the constant-expansion
phase, occurs at 8.5 Gyr ago. Also, at that time the cosmic radiation 
temperature was 146K. Recent
observations of distant supernovae imply, in defiance of expectations, that 
the Universe's growth is accelerating, contrary to what has always been 
assumed, that the expansion is slowing down due to gravity. Our theory 
confirms these recent experimental results by showing that the Universe now is 
definitely in a stage of accelerating expansion. The theory predicts also that 
now there is a positive pressure, $p=0.034g/cm^2$, in the Universe. Although 
the theory has no
cosmological constant, we extract from it its equivalence and show that 
$\Lambda=1.934\times 10^{-35}s^{-2}$. This value of 
$\Lambda$ is in excellent agreement with the measurements obtained by the
{\it High-Z Supernova Team} and the {\it Supernova Cosmology Project}. It is 
also shown that the three-dimensional space of the Universe is Euclidean, as 
the Boomerang experiment shows. Comparison with general relativity theory is
finally made and it is shown that the classical experiments as well as the 
gravitational radiation prediction follow from the present theory, too.
\end{abstract} 
\vspace{0.5cm}
\begin{center}{PACS numbers: 04.50.+h, 11.25.Mj., 11.27.+d, 98.80Cq}
\end{center}
\vspace{2cm}
$^\star$Email: carmelim@bgumail.bgu.ac.il
\newpage  
\section{Preliminaries}
As in classical general relativity we start our discussion in flat 
spacevelocity which
will then be generalized to curved space. 

The flat-spacevelocity cosmological metric is given by 
$$ds^2=\tau^2dv^2-\left(dx^2+dy^2+dz^2\right).\eqno(1)$$
Here $\tau$ is Hubble's time, the inverse of Hubble's constant, as given by
measurements in the limit of zero distances and thus zero gravity. As such, 
$\tau$ is a constant, in fact a universal constant (its
numerical value is given in Section 8, $\tau=12.486$Gyr). Its role 
in cosmology theory resembles that of $c$, the speed of light in vacuum, in 
ordinary special relativity. The velocity $v$ is used here in the sense of 
cosmology, as in Hubble's law, and is usually not the time-derivative of the 
distance.

The Universe expansion is obtained from the metric (1) as a null condition,
$ds=0$. Using spherical coordinates $r$, $\theta$, $\phi$ for the metric 
(1), and the fact that the Universe is spherically symmetric ($d\theta=
d\phi=0$), the null condition then yields $dr/dv=\tau$, or upon integration
and using appropriate initial conditions, gives $r=\tau v$ or $v=H_0r$, i.e.
the Hubble law in the zero-gravity limit.        

Based on the metric (1) a cosmological special relativity 
(CSR) was presented in the text [1] (see Chapter 2). In this theory the receding 
velocities of galaxies and the
distances between them in the Hubble expansion are united into a 
four-dimensional pseudo-Euclidean manifold, similarly to space and time in
ordinary special relativity. The Hubble law is assumed and is written in an
invariant way that enables one to derive a four-dimensional transformation 
which is similar to the Lorentz transformation. The parameter in the new 
transformation is the ratio between the cosmic time to $\tau$ (in which the 
cosmic time is measured backward with respect to the present time). 
Accordingly, the new transformation relates physical quantities at different 
cosmic times in the limit of weak or negligible gravitation. 

The transformation between the four variables $x$, $y$, $z$, $v$ and $x'$, 
$y'$, $z'$, $v'$ (assuming $y'=y$ and $z'=z$) is given by
$$x'=\frac{x-tv}{\sqrt{1-t^2/{\tau}^2}}, \ \ \ \
v'=\frac{v-tx/{\tau}^2}{\sqrt{1-t^2/{\tau}^2}}, \ \ \ \ 
y'=y, \ z'=z.\eqno(2)$$   
Equations (2) are the {\it cosmological transformation} and very much resemble 
the
well-known Lorentz transformation. In CSR it is the relative cosmic time which
takes the role of the relative velocity in Einstein's special relativity. The
transformation (2) leaves invariant the Hubble time $\tau$, just as the
Lorentz transformation leaves invariant the speed of light in vacuum $c$. 

\section{Cosmology in spacevelocity}
A cosmological general theory of relativity, suitable for the large-scale 
structure of the Universe, was subsequently developed [2-5]. In the 
framework of cosmological general relativity (CGR) gravitation is described
by a curved four-dimensional Riemannian spacevelocity. CGR incorporates the
Hubble constant $\tau$ at the outset. The Hubble law is assumed in CGR as a
fundamental law. CGR, in essence, extends Hubble's law so as to incorporate 
gravitation in it; it is actually a {\it distribution theory} that relates distances 
and velocities between galaxies. The theory involves only measured quantities
and it takes a picture of the Universe as it is at any moment. The following 
is a brief review of CGR as was originally given by the author in 1996 in Ref. 
2.

The foundations of any gravitational theory are based on the principle of
equivalence and the principle of general covariance [6]. These two principles lead immediately
to the realization that gravitation should be described by a four-dimensional
curved spacetime, in our theory spacevelocity, and that the field equations 
and the equations of motion should be written in a generally covariant form.
Hence these principles were adopted in CGR also. Use is made in a 
four-dimensional Riemannian manifold with a metric $g_{\mu\nu}$ and a line 
element $ds^2=g_{\mu\nu}dx^\mu dx^\nu$. The difference from Einstein's general
relativity is that our coordinates are: $x^0$ is a velocitylike coordinate 
(rather than a timelike coordinate), thus $x^0=\tau v$ where $\tau$ is the
Hubble time in the zero-gravity limit and $v$ the velocity. The coordinate 
$x^0=\tau v$ is the 
comparable to $x^0=ct$ where $c$ is the speed of light and $t$ is the time in
ordinary general relativity. The other three coordinates $x^k$, $k=1,2,3$, are
spacelike, just as in general relativity theory.

An immediate consequence of the above choice of coordinates is that the null
condition $ds=0$ describes the expansion of the Universe in the curved 
spacevelocity (generalized Hubble's law with gravitation) as compared to the 
propagation of light in the curved spacetime in general relativity. This means
one solves the field equations (to be given in the sequel) for the metric 
tensor, then from the null condition $ds=0$ one obtains immedialety the 
dependence of the relative distances between the galaxies on their relative
velocities.

As usual in gravitational theories, one equates geometry to physics. The first 
is expressed by means of a combination of the Ricci tensor and the Ricci
scalar, and follows to be naturally either the Ricci trace-free tensor or the
Einstein tensor. The Ricci trace-free tensor does not fit gravitation in 
general, and
the Einstein tensor is a natural candidate. The physical part is expressed by
the energy-momentum tensor which now has a different physical meaning from 
that in Einstein's theory. More important, the coupling constant that relates 
geometry to physics is now also {\it different}. 

Accordingly the field equations are
$$G_{\mu\nu}=R_{\mu\nu}-\frac{1}{2}g_{\mu\nu}R=\kappa T_{\mu\nu},\eqno(3)$$
exactly as in Einstein's theory, with $\kappa$ given by
$\kappa=8\pi k/\tau^4$, (in general relativity it is given by $8\pi G/c^4$),
where $k$ is given by $k=G\tau^2/c^2$, with $G$ being Newton's gravitational
constant, and $\tau$ the Hubble constant time. When the equations of motion 
will be written in terms of velocity instead of time, the constant $k$ will
replace $G$. Using the above equations one then has $\kappa=8\pi G/c^2\tau^2$.

The energy-momentum tensor $T^{\mu\nu}$ is constructed, along the lines of
general relativity  theory, with the speed of light being replaced by the
Hubble constant time. If $\rho$ is the average mass density of the Universe,
then it will be assumed that $T^{\mu\nu}=\rho u^\mu u^\nu,$ where $u^\mu=
dx^\mu/ds$ is the four-velocity.
In general relativity theory one takes $T_0^0=\rho$. In Newtonian gravity one
has the Poisson equation $\nabla^2\phi=4\pi G\rho$. At points where $\rho=0$
one solves the vacuum Einstein field equations in general relativity and the 
Laplace equation 
$\nabla^2\phi=0$ in Newtonian gravity. In both theories a null (zero) solution
is allowed as a trivial case. In cosmology, however, there exists no situation
at which $\rho$ can be zero because the Universe is filled with matter. In
order to be able to have zero on the right-hand side of Eq. (3) one takes 
$T_0^0$ not as equal to $\rho$, but to $\rho_{eff}=\rho-\rho_c$, where 
$\rho_c$ is the critical mass density, 
a {\it constant} in CGR given by $\rho_c=3/8\pi G\tau^2$, whose value is 
$\rho_c\approx 10^{-29}g/cm^3$, a few hydrogen atoms per cubic meter. 
Accordingly one takes
$$T^{\mu\nu}=\rho_{eff}u^\mu u^\nu;\mbox{\hspace{5mm}}\rho_{eff}=\rho-\rho_c
\eqno(4)$$
for the energy-momentum tensor.

In the next sections we apply CGR to obtain the accelerating expanding 
Universe and related subjects.
\section{Gravitational field equations}
In the four-dimensional spacevelocity the spherically symmetric metric is 
given by
$$ds^2=\tau^2dv^2-e^\mu dr^2-R^2\left(d\theta^2+\sin^2\theta d\phi^2\right),
\eqno(5)$$
where $\mu$ and $R$ are functions of $v$ and $r$ alone, and comoving 
coordinates $x^\mu=(x^0,x^1,x^2,x^3)=(\tau v,r,\theta,\phi)$ have been used. 
With the above choice of coordinates, the zero-component of the geodesic
equation becomes an identity, and since $r$, $\theta$ and $\phi$ are constants
along the geodesics, one has $dx^0=ds$ and therefore
$$u^\alpha=u_\alpha=\left(1,0,0,0\right).\eqno(6)$$
The metric (5) shows that the area of the sphere $r=constant$ is given by
$4\pi R^2$ and that $R$ should satisfy $R'=\partial R/\partial r>0$. The
possibility that $R'=0$ at a point $r_0$ is excluded since it would
allow the lines $r=constants$ at the neighboring points $r_0$ and $r_0+dr$ to
coincide at $r_0$, thus creating a caustic surface at which the comoving 
coordinates break down.

As has been shown in the previous sections the Universe expands by the null
condition $ds=0$, and if the expansion is spherically symmetric one has
$d\theta=d\phi=0$. The metric (5) then yields
$$\tau^2 dv^2-e^\mu dr^2=0,\eqno(7)$$
thus
$$\frac{dr}{dv}=\tau e^{-\mu/2}.\eqno(8)$$
This is the differential equation that determines the Universe expansion. In
the following we solve the gravitational field equations in order to find out
the function $\mu\left(r.v\right)$.

The gravitational field equations (3), written in the form
$$R_{\mu\nu}=\kappa\left(T_{\mu\nu}-\frac{1}{2}g_{\mu\nu}T\right),\eqno(9)$$
where 
$$T_{\mu\nu}=\rho_{eff}u_\mu u_\nu+p\left(u_\mu u_\nu-g_{\mu\nu}\right), 
\eqno(10)$$
with $\rho_{eff}=\rho-\rho_c$ and $T=T_{\mu\nu}g^{\mu\nu}$, are now solved.
Using Eq. (6) one finds that the only nonvanishing components of $T_{\mu\nu}$ 
are $T_{00}=\tau^2\rho_{eff}$, $T_{11}=c^{-1}\tau pe^\mu$, $T_{22}=c^{-1}\tau 
pR^2$ and $T_{33}=c^{-1}\tau pR^2\sin^2\theta$, and that $T=\tau^2\rho_{eff}-
3c^{-1}\tau p$.

The only nonvanishing components of the Ricci tensor yield (dots and primes 
denote differentiation with respect to $v$ and $r$, respectively), using Eq. 
(9), the following field equations:
$$R_{00}=-\frac{1}{2}\ddot{\mu}-\frac{2}{R}\ddot{R}-\frac{1}{4}\dot{\mu}^2=
\frac{\kappa}{2}\left(\tau^2\rho_{eff}+3c^{-1}\tau p\right),\eqno(11a)$$
$$R_{01}=\frac{1}{R}R'\dot{\mu}-\frac{2}{R}\dot{R}'=0,\eqno(11b)$$
$$R_{11}=e^\mu\left(\frac{1}{2}\ddot{\mu}+\frac{1}{4}\dot{\mu}^2+\frac{1}{R}
\dot{\mu}\dot{R}\right)+\frac{1}{R}\left(\mu'R'-2R''\right)$$
$$=\frac{\kappa}{2}e^\mu\left(\tau^2\rho_{eff}-c^{-1}\tau p\right),
\eqno(11c)$$
$$R_{22}=R\ddot{R}+\frac{1}{2}R\dot{R}\dot{\mu}+\dot{R}^2+1-e^{-\mu}\left(
RR''-\frac{1}{2}RR'\mu'+R'^2\right)$$
$$=\frac{\kappa}{2}R^2\left(\tau^2\rho_{eff}-c^{-1}\tau p\right),\eqno(11d)$$
$$R_{33}=\sin^2\theta R_{22}=\frac{\kappa}{2}R^2\sin^2\theta\left(\tau^2
\rho_{eff}-c^{-1}\tau p\right).\eqno(11e)$$

The field equations obtained for the components 00, 01, 11, and 22 (the 33 
component contributes no new information) are given by
$$-\ddot{\mu}-\frac{4}{R}\ddot{R}-\frac{1}{2}\dot{\mu}^2=\kappa\left(
\tau^2\rho_{eff}+3c^{-1}\tau p\right),
\eqno(12)$$
$$2\dot{R}'-R'\dot{\mu}=0, \eqno(13)$$
$$\ddot{\mu}+\frac{1}{2}\dot{\mu}^2+\frac{2}{R}\dot{R}\dot{\mu}+e^{-\mu}\left(
\frac{2}{R}R'\mu'-\frac{4}{R}R''\right)=\kappa\left(\tau^2\rho_{eff}-c^{-1}\tau 
p\right) \eqno(14)$$
$$\frac{2}{R}\ddot{R}+2\left(\frac{\dot{R}}{R}\right)^2+\frac{1}{R}\dot{R}
\dot{\mu}+\frac{2}{R^2}+e^{-\mu}\left[\frac{1}{R}R'\mu'-2\left(\frac{R'}{R}
\right)^2-\frac{2}{R}R''\right]$$
$$=\kappa\left(\tau^2\rho_{eff}-c^{-1}\tau p\right).\eqno(15)$$
It is convenient to eliminate the term with the second velocity-derivative of
$\mu$ from the above equations. This can easily be done, and combinations of 
Eqs. (12)--(15) then give the following set of three independent field 
equations:
$$e^\mu\left(2R\ddot{R}+\dot{R}^2+1\right)-R'^2=-\kappa\tau c^{-1} e^\mu R^2p,
\eqno(16)$$
$$2\dot{R}'-R'\dot{\mu}=0, \eqno(17)$$
$$e^{-\mu}\left[\frac{1}{R}R'\mu'-\left(\frac{R'}{R}\right)^2-\frac{2}{R}R''
\right]+\frac{1}{R}\dot{R}\dot{\mu}+\left(\frac{\dot{R}}{R}\right)^2+
\frac{1}{R^2}$$
$$=\kappa\tau^2\rho_{eff}, \eqno(18)$$
other equations being trivial combinations of (16)--(18).
\section{Solution of the field equations}
The solution of Eq. (17) satisfying the condition $R'>0$ is given by
$$e^\mu=\frac{R'^2}{1+f\left(r\right)},\eqno(19)$$
where $f\left(r\right)$ is an arbitrary function of the coordinate $r$ and 
satisfies the
condition $f\left(r\right)+1>0$. Substituting (19) in the other two field
equations (16) and (18) then gives
$$2R\ddot{R}+\dot{R}^2-f=-\kappa c^{-1}\tau R^2p, \eqno(20)$$
$$\frac{1}{RR'}\left(2\dot{R}\dot{R'}-f'\right)+\frac{1}{R^2}\left(\dot{R}^2-f
\right)=\kappa\tau^2\rho_{eff},\eqno(21)$$
respectively.

The simplest solution of the above two equations, which satisfies the 
condition $R'=1>0$, is given by
$$R=r.\eqno(22)$$
Using Eq. (22) in Eqs. (20) and (21) gives
$$f\left(r\right)=\kappa c^{-1}\tau pr^2,\eqno(23)$$
and
$$f'+\frac{f}{r}=-\kappa\tau^2\rho_{eff}r,\eqno(24)$$
respectively. The solution of Eq. (24) is the sum of the solutions of the
homogeneous equation
$$f'+\frac{f}{r}=0,\eqno(25)$$
and a particular solution of Eq. (24). These are given by
$$f_1=-\frac{2Gm}{c^2r},\eqno(26)$$
and 
$$f_2=-\frac{\kappa}{3}\tau^2\rho_{eff}r^2.\eqno(27)$$

The solution $f_1$ represents a particle at the origin of coordinates and as
such is not relevant to our problem. We take, accordingly, $f_2$ as the 
general solution,
$$f\left(r\right)=-\frac{\kappa}{3}\tau^2\rho_{eff}r^2=-\frac{\kappa}{3}\tau^2
\left(\rho-\rho_c\right)r^2$$
$$=-\frac{\kappa}{3}\tau^2\rho_c\left(
\frac{\rho}{\rho_c}-1\right)r^2.\eqno(28)$$
Using the values of $\kappa=8\pi G/c^2\tau^2$ and $\rho_c=3/8\pi G\tau^2$, we
obtain
$$f\left(r\right)=\frac{1-\Omega_m}{c^2\tau^2}r^2,\eqno(29)$$
where $\Omega_m=\rho/\rho_c$.

The two solutions given by Eqs. (23) and (29) for $f(r)$ can now be 
equated, giving
$$p=\frac{1-\Omega_m}{\kappa c\tau^3}=\frac{c}{\tau}\frac{1-\Omega_m}{8\pi G}
=4.544\left(1-\Omega_m\right)\times 10^{-2} g/cm^2.\eqno(30)$$
Furthermore, from Eqs. (19) and (22) we find that 
$$e^{-\mu}=1+f\left(r\right)=1+\tau c^{-1}\kappa pr^2=1+
\frac{1-\Omega_m}{c^2\tau^2}r^2.\eqno(31)$$

It  will be recalled that the Universe expansion is determined by Eq. (8),
$dr/dv=\tau e^{-\mu/2}$. The only thing that is left to be determined is the
signs of $(1-\Omega_m)$ or the pressure $p$.

Thus we have
$$\frac{dr}{dv}=\tau\sqrt{1+\kappa\tau c^{-1}pr^2}=\tau\sqrt{1+
\frac{1-\Omega_m}{c^2\tau^2}r^2}.\eqno(32)$$
For simplicity we confine ourselves to the linear approximation, thus Eq. 
(32) yields
$$\frac{dr}{dv}=\tau\left(1+\frac{\kappa}{2}\tau c^{-1}pr^2\right)=
\tau\left[1+\frac{1-\Omega_m}{2c^2\tau^2}r^2\right].\eqno(33)$$
\section{Classification of universes}
The second term in the square bracket in the above equation represents the
deviation due to gravity from the standard Hubble law. For without that term,
Eq. (33) reduces to $dr/dv=\tau$, thus $r=\tau v+const$. The constant can be 
taken zero if one assumes, as usual, that at $r=0$ the velocity should also
vanish. Thus $r=\tau v$, or $v=H_0r$ (since $H_0\approx 1/\tau$). Accordingly, 
the equation of motion (33) describes the expansion of the Universe when 
$\Omega_m=1$, namely when $\rho=\rho_c$. The equation then coincides with the 
standard Hubble law.

The equation of motion (33) can easily be integrated exactly by the
substitions
$$\sin\chi =\sqrt{\frac{\left(\Omega_m-1\right)}{2}}\frac{r}{2c\tau};\hspace{5mm}
\Omega_m>1,\eqno(34a)$$
$$\sinh\chi =\sqrt{\frac{\left(1-\Omega_m\right)}{2}}\frac{r}{2c\tau};
\hspace{5mm}\Omega_m<1.\eqno(34b)$$
One then obtains, using Eqs. (33) and (34),
$$dv=cd\chi/\left(\Omega_m-1\right)^{1/2}\cos\chi;\hspace{5mm}\Omega_m>1,\eqno(35a)$$
$$dv=cd\chi/\left(1-\Omega_m\right)^{1/2}\cosh\chi;\hspace{5mm}\Omega_m<1.\eqno(35b)$$

We give below the exact solutions for the expansion of the Universe for each 
of the cases, $\Omega_m>1$ and $\Omega_m<1$. As will be seen, the case of 
$\Omega_m=1$ can be obtained at the limit $\Omega_m\rightarrow 1$ from both cases.

{\bf The case $\Omega_m>1$.}
From Eq. (35a) we have
$$\int dv=\frac{c}{\sqrt{\left(\Omega_m-1\right)/2}}\int\frac{d\chi}{\cos\chi},
\eqno(36)$$
where $\sin\chi=r/a$, and $a=c\tau\sqrt{\left(\Omega_m-1\right)/2}$. A simple 
calculation gives [7]
$$\int\frac{d\chi}{\cos\chi}=\ln\left|\frac{1+\sin\chi}{\cos\chi}\right|.
\eqno(37)$$
A straightforward calculation then gives
$$v=\frac{a}{2\tau}\ln\left|\frac{1+r/a}{1-r/a}\right|.\eqno(38)$$
As is seen, when $r\rightarrow 0$ then $v\rightarrow 0$ and using the
L'Hospital lemma, $v\rightarrow r/\tau$ as $a\rightarrow 0$ (and thus
$\Omega_m\rightarrow 1$).

{\bf The case $\Omega_m<1$.} 
From Eq. (35b) we now have
$$\int dv=\frac{c}{\sqrt{\left(1-\Omega_m\right)/2}}\int\frac{d\chi}{\cosh\chi},
\eqno(39)$$ 
where $\sinh\chi=r/b$, and $b=c\tau\sqrt{\left(1-\Omega_m\right)/2}$. A 
straightforward calculation then gives [7]
$$\int\frac{d\chi}{\cosh\chi}=\arctan e^\chi.\eqno(40)$$  
We then obtain
$$\cosh\chi=\sqrt{1+\frac{r^2}{b^2}},\eqno(41)$$
$$e^\chi=\sinh\chi+\cosh\chi=\frac{r}{b}+\sqrt{1+\frac{r^2}{b^2}}.\eqno(42)$$
Equations (39) and (40) now give
$$v=\frac{2c}{\sqrt{\left(1-\Omega_m\right)/2}}\arctan e^\chi +K,\eqno(43)$$
where $K$ is an integration constant which is determined by the requirement
that at $r=0$ then $v$ should be zero. We obtain
$$K=-\pi c/2\sqrt{\left(1-\Omega_m\right)/2},\eqno(44)$$
and thus
$$v=\frac{2c}{\sqrt{\left(1-\Omega_m\right)/2}}\left(\arctan e^\chi-
\frac{\pi}{4}\right).\eqno(45)$$
A straightforward calculation then gives
$$v=\frac{b}{\tau}\left\{2\arctan\left(\frac{r}{b}+\sqrt{1+\frac{r^2}{b^2}}
\right)-\frac{\pi}{2}\right\}.\eqno(46)$$
As for the case $\Omega_m>1$ one finds that $v\rightarrow 0$ when $r\rightarrow 
0$, and again, using L'Hospital lemma, $r=\tau v$ when $b\rightarrow 0$ 
(and thus $\Omega_m\rightarrow 1$).

\section{Physical meaning}
To see the physical meaning of these solutions, however, one does not need the
exact solutions. Rather, it is enough to write down the solutions in the
lowest approximation in $\tau^{-1}$. One obtains, by differentiating Eq. 
(33) with respect to $v$, for $\Omega_m>1$,
$$d^2r/dv^2=-kr;\mbox{\hspace{10mm}}k=\frac{\left(\Omega_m-1\right)}{2c^2},\eqno(47)$$
the solution of which is 
$$r\left(v\right)=A\sin\alpha\frac{v}{c}+B\cos\alpha\frac{v}{c},\eqno(48)$$
where $\alpha^2=(\Omega_m-1)/2$ and $A$ and $B$ are constants. The latter can be
determined by the initial condition $r\left(0\right)=0=B$ and $dr\left(0
\right)/dv=\tau=A\alpha/c$, thus
$$r\left(v\right)=\frac{c\tau}{\alpha}\sin\alpha\frac{v}{c}.\eqno(49)$$
This is obviously a closed Universe, and presents a decelerating expansion.

For $\Omega_m<1$ we have
$$d^2r/dv^2=\frac{\left(1-\Omega_m\right)r}{2c^2},\eqno(50)$$
whose solution, using the same initial conditions, is
$$r\left(v\right)=\frac{c\tau}{\beta}\sinh\beta\frac{v}{c},\eqno(51)$$
where $\beta^2=(1-\Omega_m)/2$. This is now an open accelerating Universe.

For $\Omega_m=1$ we have, of course, $r=\tau v$.
\section{The accelerating universe}
We finally determine which of the three cases of expansion is the one at 
present epoch of time. To this end we have to write the solutions (49) and 
(51) in ordinary Hubble's law form $v=H_0r$. Expanding Eqs. (49) and 
(51) into power series in $v/c$ and keeping terms up to the second order, we
obtain
$$r=\tau v\left(1-\alpha^2v^2/6c^2\right), \eqno(52a)$$
$$r=\tau v\left(1+\beta^2v^2/6c^2\right), \eqno(52b)$$
for $\Omega_m>1$ and $\Omega_m<1$, respectively. Using now the expressions for 
$\alpha$ and $\beta$, Eqs. (52) then reduce into the single equation
$$r=\tau v\left[1+\left(1-\Omega_m\right)v^2/6c^2\right].\eqno(53)$$
Inverting now this equation by writing it as $v=H_0r$, we obtain in the lowest
approximation
$$H_0=h\left[1-\left(1-\Omega_m\right)v^2/6c^2\right],\eqno(54)$$
where $h=\tau^{-1}$. To the same approximation one also obtains
$$H_0=h\left[1-\left(1-\Omega_m\right)z^2/6\right]=h\left[1-\left(1-\Omega_m
\right)r^2/6c^2\tau^2\right],\eqno(55)$$
where $z$ is the redshift parameter.
As is seen, and it is confirmed by experiments, $H_0$ depends on the distance 
it is being measured; it has physical meaning only at the zero-distance limit,
namely when measured {\it locally}, in which case it becomes $h=1/\tau$.

It follows that the measured value of $H_0$ depends on the ``short"
and ``long" distance scales [8]. The farther the distance $H_0$ is being 
measured, the lower the value for $H_0$ is obtained. By Eq. (55) this is
possible only when $\Omega_m<1$, namely when the Universe is accelerating. By
Eq. (30) we also find that the pressure is positive. 

The possibility that the Universe expansion is accelerating was first 
predicted using CGR by the author in 1996 [2] before the supernovae 
experiments results became known.

It  will be noted that the constant expansion is just a transition stage 
between the decelerating and the accelerating expansions as the Universe
evolves toward its present situation.

Figure 1 describes the Hubble diagram of the above solutions for the three 
types of expansion for values of $\Omega_m$ from 100 to 0.245. The figure
describes the three-phase evolution of the Universe. Curves (1)-(5) represent
the stages of {\it decelerating expansion} according to Eq. (49). As the density 
of matter $\rho$ decreases, the Universe goes over from the lower curves to 
the upper ones, but it does not have enough time to close up to a big crunch.
The Universe subsequently goes over to curve (6) with $\Omega_m=1$, at which time
it has a constant expansion for a fraction of a second. This then followed
by going to the upper curves (7) and (8) with $\Omega_m<1$, where the Universe
expands with {\it acceleration} according to Eq. (51). Curve no. 8 fits
the present situation of the Universe. For curves (1)-(4) in the diagram we
use the cutoff when the curves were at their maximum. In Table 1 we present 
the cosmic times with respect to the big bang, the cosmic radiation
temperature and the pressure for each of the curves in Fig. 1.

Figures 2 and 3 show the Hubble diagrams for the distance-redshift 
relationship predicted by the theory for the accelerating expanding Universe 
at the present time, and Figures 4 and 5 show the experimental results.

Our estimate for $h$, based on published data, is $h\approx 80$ km/sec-Mpc.
Assuming $\tau^{-1}\approx 80$ km/sec-Mpc, Eq. (55) then gives
$$H_0=h\left[1-1.3\times 10^{-4}\left(1-\Omega_m\right)r^2\right],\eqno(56)$$
where $r$ is in Mpc. A computer best-fit can then fix both $h$ and $\Omega_m$.

To summarize, a theory of cosmology has been presented in which the dynamical
variables are those of Hubble, i.e. distances and velocities. The theory
descirbes the Universe as having a three-phase evolution with a decelerating 
expansion, followed by a constant and an accelerating expansion, and it
predicts that the Universe is now in the latter phase. As the density of 
matter decreases, while the Universe is at the decelerating phase, it does not
have enough time to close up to a big crunch. Rather, it goes to the 
constant-expansion phase, and then to the accelerating stage. As we have seen,
the equation obtained for the Universe expansion, Eq. (51), is very simple.

\section{Theory versus experiment}
The Einstein gravitational 
field equations  
with the added cosmological term are [9]:
$$R_{\mu\nu}-\frac{1}{2}g_{\mu\nu}R+\Lambda g_{\mu\nu}=\kappa T_{\mu\nu},
\eqno(57)$$
where $\Lambda$ is the cosmological constant, the value of which is supposed to
be determined by experiment. In Eq. (57) $R_{\mu\nu}$ and $R$ are the Ricci 
tensor and scalar, respectively, $\kappa=8\pi G$, where $G$ is Newton's constant
and the speed of light is taken as unity.

Recently the two groups (the {\it Supernovae Cosmology Project} and the {\it 
High-Z Supernova Team}) concluded that the expansion of the Universe is 
accelerating [10-16]. The two groups had discovered and measured moderately 
high redshift ($0.3<z<0.9$) supernovae, and found that they were fainter than
what one would expect them to be if the cosmos expansion were slowing down or
constant. Both teams obtained
$$\Omega_m\approx 0.3,\hspace{5mm} \Omega_\Lambda\approx 0.7,\eqno(58)$$
and ruled out the traditional ($\Omega_m$, $\Omega_\Lambda$)=(1, 0)
Universe. Their value of the density parameter $\Omega_\Lambda$ corresponds to
a cosmological constant that is small but, nevertheless, nonzero and positive,
$$\Lambda\approx 10^{-52}\mbox{\rm m}^{-2}\approx 10^{-35}\mbox{\rm s}^{-2}.
\eqno(59)$$

In previous sections a four-dimensional cosmological theory (CGR) was 
presented. Although the theory has no cosmological constant, it predicts that 
the Universe accelerates and hence it has the equivalence of a positive 
cosmological constant in 
Einstein's general relativity. In the framework of this theory (see Section 
2) the 
zero-zero component of the field equations (3) is written as
$$R_0^0-\frac{1}{2}\delta_0^0R=\kappa\rho_{eff}=\kappa\left(\rho-\rho_c
\right),\eqno(60)$$
where $\rho_c=3/\kappa\tau^2$ is the critical mass density
and $\tau$ is Hubble's time in the zero-gravity limit.

Comparing Eq. (60) with the zero-zero component of Eq. (57), one obtains 
the expression for the cosmological constant of general relativity,
$$\Lambda=\kappa\rho_c=3/\tau^2.\eqno(61)$$

To find out the numerical value of $\tau$ we use the relationship between
$h=\tau^{-1}$ and $H_0$ given by Eq. (55) (CR denote values according to
Cosmological Relativity): 
$$H_0=h\left[1-\left(1-\Omega_m^{CR}\right)z^2/6\right],\eqno(62)$$
where $z=v/c$ is the redshift and $\Omega_m^{CR}=\rho_m/\rho_c$ with 
$\rho_c=3h^2/8\pi G$. (Notice that our $\rho_c=1.194\times 10^{-29}g/cm^3$ is different from 
the standard $\rho_c$ defined with $H_0$.) The redshift parameter $z$ 
determines the distance at which $H_0$ is measured. We choose $z=1$ and take 
for 
$$\Omega_m^{CR}=0.245, \eqno(63)$$
its value at the present time (see Table 1) (corresponds to 0.32 in the 
standard theory), Eq. (62) then gives
$$H_0=0.874h.\eqno(64)$$
At the value $z=1$ the corresponding Hubble parameter $H_0$ according to the 
latest results from HST can be taken [17] as $H_0=70$km/s-Mpc, thus 
$h=(70/0.874)$km/s-Mpc, or
$$h=80.092\mbox{\rm km/s-Mpc},\eqno(65)$$
and 
$$\tau=12.486 Gyr=3.938\times 10^{17}s.\eqno(66)$$

What is left is to find the value of $\Omega_\Lambda^{CR}$. We have 
$\Omega_\Lambda^{CR}=\rho_c^{ST}/\rho_c$, where $\rho_c^{ST}=3H_0^2/8\pi 
G$ and $\rho_c=3h^2/8\pi G$. Thus $\Omega_\Lambda^{CR}=(H_0/h)^2=0.874^2$,
or
$$\Omega_\Lambda^{CR}=0.764.\eqno(67)$$
As is seen from Eqs. (63) and (67) one has 
$$\Omega_T=\Omega_m^{CR}+\Omega_\Lambda^{CR}=0.245+0.764=1.009\approx 1,
\eqno(68)$$
which means the Universe is Euclidean.

As a final result we calculate the cosmological constant according to 
Eq. (61). One obtains
$$\Lambda=3/\tau^2=1.934\times 10^{-35}s^{-2}.\eqno(69)$$

Our results confirm those of the supernovae experiments and indicate on the
existance of the dark energy as has recently received confirmation from the
Boomerang cosmic microwave background experiment [18,19], which showed that 
the Universe is Euclidean.
\section{Some remarks}
In this paper the cosmological general relativity, a relativistic theory in
spacevelocity, has been presented and applied to the problem of the expansion 
of the Universe. The theory, which predicts a positive pressure for the 
Universe now, describes the Universe as having a three-phase
evolution: decelerating, constant and accelerating expansion, but it is now in
the latter stage. Furthermore, the cosmological constant that was
extracted from the theory agrees with the experimental result. Finally, it has
also been shown that the three-dimensional spatial space of the Universe is
Euclidean, again in agreement with observations. 

Recently [20,21], more confirmation to the Universe accelerating expansion 
came from 
the most distant supernova, SN 1997ff, that was recorded by the Hubble Space
Telescope. As has been pointed out before, if we look back far enough, we 
should find a decelerating expansion (curves 1-5 in Figure 1). Beyond $z=1$
one should see an earlier time when the mass density was dominant. The 
measurements obtained from SN 1997ff's redshift and brightness provide a 
direct proof for the transition from past decelerating to present 
accelerating expansion (see Figures 6 and 7). The measurements also exclude 
the possibility that
the acceleration of the Universe is not real but is due to other astrophysical 
effects such as dust.

Table 2 gives some of the cosmological parameters obtained here and in the
standard theory.
\section{Comparison with general relativity}
In order to compare the present theory with general relativity, we now add the 
time coordinate. We then have a time-space-velocity Universe with two 
time-like and three space-like coordinates, with signature $(+---+)$. We will
be concerned with the classical experiments of general relativity and the 
gravitational waves predicted by that theory. In the following we show that
all these results are also obtained from the present theory. To this end we proceed as
follows. 

We first find the cosmological-equivalent of the Schwarzschild 
spherically-symmetric solution in cosmology. It will be useful to change 
variables from the classical Schwarzschild metric to new variables as follows:
$$\sin^2\chi=r_s/r,\mbox{\hspace{5mm}}
dr=-2r_s\sin^{-3}\chi\cos\chi d\chi,\eqno(70)$$
where $r_s=2GM/c^2$ is the Schwarzschild radius. We also change the time 
coordinate $cdt=r_sd\eta$, thus $\eta$ is a time parameter. The classical
Schwarzschild solution will thus have the following form:   
$$ds^2=r_s^2\left[\cos^2\chi d\eta^2-4\sin^{-6}\chi d\chi^2-\sin^{-4}\chi
\left(d\theta^2+\sin^2\theta d\phi^2\right)\right].\eqno(71)$$
So far this is just the classical spherically symmetric solution of the 
Einstein field equations in four dimensions, though written in new variables.
The non-zero Christoffel symbols are given by
$$\Gamma^0_{01}=-\sin\chi\cos^{-1}\chi,\mbox{\hspace{5mm}}
\Gamma^1_{00}=-\frac{1}{4}\sin^7\chi\cos\chi,$$
$$\Gamma^1_{11}=-3\sin^{-1}\chi\cos\chi,\mbox{\hspace{5mm}}
\Gamma^1_{22}=\frac{1}{2}\sin\chi\cos\chi,$$
$$\Gamma^1_{33}=\frac{1}{2}\sin\chi\cos\chi\sin^2\theta,\mbox{\hspace{5mm}}
\Gamma^2_{12}=-2\sin^{-1}\chi\cos\chi,\eqno(72)$$
$$\Gamma^2_{33}=-\sin\theta\cos\theta,\mbox{\hspace{5mm}}
\Gamma^3_{13}=-2\sin^{-1}\chi\cos\chi,\mbox{\hspace{5mm}}
\Gamma^3_{23}=\sin^{-1}\theta\cos\theta.$$
It is very lengthy, but one can verify that all components of the Ricci tensor
$R_{\alpha\beta}$ are equal to zero identically. 

We now extend this solution
to cosmology. In order to conform with the standard notation, the zero 
component will be chosen as the time parameter, followed by the three 
space-like coordinates and then the fourth coordinate representing the 
velocity $\tau dv$. We will make one more change by choosing $\tau dv=r_sdu$,
thus $u$ is the velocity parameter. The simplest way to have a cosmological 
solution of the Einstein field equation is using the so-called co-moving
coordinates in which:
$$ds^2=r_s^2\left[\cos^2\chi d\eta^2-4\sin^{-6}\chi d\chi^2-\sin^{-4}\chi
\left(d\theta^2+\sin^2\theta d\phi^2\right)+du^2\right].\eqno(73)$$
The coordinates are now $x^0=\eta$, $x^1=\chi$, $x^2=\theta$, $x^3=\phi$, and 
$x^4=u$, and $r_s$ is now a function of the velocity $u$, $r_s=r_s(u)$ to be
determined by the Einstein field equations in five dimensions. Accordingly we
have the following form for the metric:
$$g_{\mu\nu}=r_s^2\left(\begin{array}{ccccc}
\cos^2\chi&&&&0\\
&-4\sin^{-6}\chi&&&\\
&&-\sin^{-4}\chi&&\\
&&&-\sin^{-4}\chi\sin^2\theta&\\
0&&&&1\\\end{array}\right),\eqno(74a)$$
$$\sqrt{-g}=2r_s^5\sin^{-7}\chi\cos\chi\sin\theta.\eqno(74b)$$

The non-zero Christoffel symbols are given by 
$$\Gamma^0_{01}=-\sin\chi\cos^{-1}\chi, \mbox{\hspace{5mm}}
\Gamma^0_{04}=\dot{r_s}r_s^{-1},$$
$$\Gamma^1_{00}=-\frac{1}{4}\sin^7\chi\cos\chi,\mbox{\hspace{5mm}}
\Gamma^1_{11}=-3\sin^{-1}\chi\cos\chi,$$
$$\Gamma^1_{14}=\dot{r_s}r_s^{-1},\mbox{\hspace{5mm}}
\Gamma^1_{22}=\frac{1}{2}\sin\chi\cos\chi,\mbox{\hspace{5mm}}
\Gamma^1_{33}=\frac{1}{2}\sin\chi\cos\chi\sin^2\theta,$$
$$\Gamma^2_{12}=-2\sin^{-1}\chi\cos\chi,\mbox{\hspace{5mm}}
\Gamma^2_{24}=\dot{r_s}r_s^{-1},\mbox{\hspace{5mm}}
\Gamma^2_{33}=-\sin\theta\cos\theta,\eqno(75)$$
$$\Gamma^3_{13}=-2\sin^{-1}\chi\cos\chi,\mbox{\hspace{5mm}}
\Gamma^3_{23}=\sin^{-1}\theta\cos\theta,\mbox{\hspace{5mm}}
\Gamma^3_{34}=\dot{r_s}r_s^{-1},$$
$$\Gamma^4_{00}=-\dot{r_s}r_s^{-1}\cos^2\chi,\mbox{\hspace{5mm}}
\Gamma^4_{11}=4\dot{r_s}r_s^{-1}\sin^{-6}\chi,\mbox{\hspace{5mm}}
\Gamma^4_{22}=\dot{r_s}r_s^{-1}\sin^{-4}\chi,$$
$$\Gamma^4_{33}=\dot{r_s}r_s^{-1}\sin^{-4}\chi\sin^2\theta,\mbox{\hspace{5mm}}
\Gamma^4_{44}=\dot{r_s}r_s^{-1},$$
where the dots denote derivatives with respect to the velocity parameter $u$.

The Ricci tensor components after a lengthy but straightforward calculation, 
are given by:
$$R_{00}=-\left(\ddot{r_s}r_s^{-1}+2\dot{r_s}^2r_s^{-2}\right)\cos^2\chi,$$
$$R_{11}=4\left(\ddot{r_s}r_s^{-1}+2\dot{r_s}^2r_s^{-2}\right)\sin^{-6}\chi,$$
$$R_{22}=\left(\ddot{r_s}r_s^{-1}+2\dot{r_s}^2r_s^{-2}\right)\sin^{-4}\chi,\eqno(76)$$
$$R_{33}=\left(\ddot{r_s}r_s^{-1}+2\dot{r_s}^2r_s^{-2}\right)\sin^{-4}\chi
\sin^2\theta,$$
$$R_{44}=-4\left(\ddot{r_s}r_s^{-1}-\dot{r_s}^2r_s^{-2}\right).$$
All other components are identically zero.

We are interested in vacuum solution of the Einstein field  equations for
the spherically symmetric metric (generalized Schwarzschild to cosmology), 
the right-hand sides of the above equations should be taken zero. A simple
calculation then shows that $\dot{r_s}=0$, $\ddot{r_s}=0$. Accordingly the
cosmological Schwarzschild metric is given by Eq. (74a) with a constant 
$r_s=2GM/c^2$. The metric (74a) can then be written, using the coordinate 
transformations (70), as 
$$g_{\mu\nu}=\left(\begin{array}{ccccc}
1-\frac{r_s}{r}&&&&0\\
&-\left(1-\frac{r_s}{r}\right)^{-1}&&&\\ 
&&-r^2&&\\
&&&-r^2\sin^2\theta&\\
0&&&&1\\\end{array}\right),\eqno(77)$$
where the coordinates are now $x^0=ct$, $x^1=r$, $x^2=\theta$, $x^3=\phi$, and
$x^4=\tau v$. 

We are now in a position to compare the present theory with 
general relativity.
\section{Gravitational redshift}
We start with the simplest experiment of the gravitational redshift. Although 
this experiment is not considered as one of the proofs of general relativity 
(it can be derived from conservation laws and Newtonian theory). 

Consider two clocks at rest at two points denoted by 1 and 2. The propagation
of light is determined by $ds$ at each point. Since at 
these points all spatial infinitesimal displacements and change in velocities 
vanish, one has $ds^2=g_{00}c^2dt^2.$ Hence at the two points 
we have
$$ds\left(1\right)=\left[g_{00}\left(1\right)\right]^{1/2}cdt,\eqno(78a)$$
$$ds\left(2\right)=\left[g_{00}\left(2\right)\right]^{1/2}cdt\eqno(78b)$$
for the proper time (see Fig. 8).

The ratio of the rates of similar clocks, located at different places in a
gravitational field, is therefore given by
$$ds\left(2\right)/ds\left(1\right)=
\left[g_{00}\left(2\right)/g_{00}\left(1\right)\right]^{1/2}.
\eqno(79)$$
The frequency $\nu_0$ of an atom located at point 1, when measured by an
observer located at point 2, is therefore given by 
$$\nu=\nu_0\left[
g_{00}\left(1\right)/g_{00}\left(2\right)\right]^{1/2}.
\eqno(80)$$

If the gravitational field is produced by a spherically symmetric mass 
distribution, then we may use the generalized Schwarzschild metric given above
to calculate the above ratio at the two points. In this case 
$g_{00}=1-2GM/c^2r$, and therefore
$$\left[g_{00}\left(1\right)/g_{00}\left(2\right)\right]^{1/2}\approx
1+\left(GM/c^2\right)\left(1/r_2-1/r_1\right)$$
to first order in $GM/c^2r$. We thus obtain
$$\Delta\nu/\nu_0=\left(\nu-\nu_0\right)/\nu_0\approx -\left(GM/c^2\right)
\left(1/r_1-1/r_2\right)$$
for the frequency shift per unit frequency. Taking now $r_1$ to be the 
observed radius of the Sun and $r_2$ the radius of the Earth's orbit around
the Sun, then we find that
$$\Delta\nu/\nu_0\approx -GM_{Sun}/c^2r_{Sun},\eqno(81)$$
where $M_{Sun}$ and $r_{Sun}$ are the mass and radius of the Sun. Accordingly
we obtain $\Delta\nu/\nu_0\approx -2.12\times 10^{-6}$ for the frequency shift
per unit frequency of the light emitted from the Sun. The calculation made
above amounts to neglecting completely the Earth's gravitational field. The 
above result is the standard gravitational redshift (also known as the 
gravitational time dilation). 
\section{Motion in a centrally symmetric gravitational field}
We assume that small test particles move along geodesics in the gravitational
field. We also assume that planets have small masses as compared with the
mass of the Sun, to the extent that they can be considered as test particles
moving in the gravitational field of the Sun. As a  result of these 
assumptions, the geodesic equation in the cosmological Schwarzschild field 
will be taken to
describe the equation of motion of a planet moving in the gravitational field
of the Sun. In fact, we do not need the exact solution of the generalized 
Schwarzschild metric (77), but just its first approximation. We obtain in the
first approximation the following expressions for the components of the
metric tensor:
$$g_{00}=1-r_s/r,\mbox{\hspace{5mm}}g_{0m}=0,\mbox{\hspace{5mm}}g_{04}=0,$$
$$g_{mn}=-\delta_{mn}-r_sx^mx^n/r^3,\mbox{\hspace{5mm}}
g_{m4}=0,\mbox{\hspace{5mm}}g_{44}=1.\eqno(82a)$$
The contravariant components of the metric tensor are consequently given, in
the same approximation, by 
$$g^{00}=1+r_s/r,\mbox{\hspace{5mm}}g^{0m}=0,\mbox{\hspace{5mm}}g^{04}=0,$$
$$g^{mn}=-\delta^{mn}+r_sx^mx^n/r^3,\mbox{\hspace{5mm}}
g^{m4}=0,\mbox{\hspace{5mm}}g^{44}=1.\eqno(82b)$$
We may indeed verify that the relation $g_{\mu\lambda}g^{\lambda\nu}=
\delta_\mu^\nu$ between the contravariant and covariant components of the
above approximate metric tensor is satisfied to orders of magnitude of the
square of $r_s/r$. A straightforward calculation then gives the following
expressions for the Christoffel symbols:
$$\Gamma^0_{0n}=-\frac{r_s}{2}\frac{\partial}{\partial x^n}\left(\frac{1}{r}
\right),$$
$$\Gamma^k_{00}=--\frac{r_s}{2}\left(1-\frac{r_s}{r}\right)
\frac{\partial}{\partial x^k}\left(\frac{1}{r}\right),\eqno(83)$$
$$\Gamma^k_{mn}=r_s\frac{x^k}{r^3}\delta_{mn}-
\frac{3}{2}r_s\frac{x^kx^mx^n}{r^5}.$$
All other components vanish.

We now use these expressions for the Christoffel symbols in the geodesic 
equation
$$\ddot{x^k}+\left(\Gamma^k_{\alpha\beta}-\Gamma^0_{\alpha\beta}\dot{x^k}
\right)\dot{x^\alpha}\dot{x^\beta}=0,\eqno(84)$$
where a dot denotes differentiation with respect to the time coordinate $x^0$.
We obtain
$$\Gamma^0_{\alpha\beta}\dot{x^\alpha}\dot{x^\beta}=\Gamma^0_{00}+
2\Gamma^0_{0n}\dot{x^n}+2\Gamma^0_{04}\dot{x^4}+
\Gamma^0_{mn}\dot{x^m}\dot{x^n}+2\Gamma^0_{m4}\dot{x^m}\dot{x^4}+
\Gamma^0_{44}\dot{x^4}\dot{x^4}$$
$$=-r_s\dot{x^n}
\frac{\partial}{\partial x^n}\left(\frac{1}{r}\right),\eqno(85a)$$
$$\Gamma^k_{\alpha\beta}\dot{x^\alpha}\dot{x^\beta}=\Gamma^k_{00}+
2\Gamma^k_{0l}\dot{x^l}+2\Gamma^k_{04}\dot{x^4}+\Gamma^k_{mn}\dot{x^m}\dot{x^n}
+2\Gamma^k_{m4}\dot{x^m}\dot{x^4}+\Gamma^k_{44}\dot{x^4}\dot{x^4}$$
$$=-\frac{r_s}{2}
\frac{\partial}{\partial x^k}\left(\frac{1}{r}\right)+
r_s\left[\frac{r_s}{2r}\frac{\partial}{\partial x^k}\left(\frac{1}{r}\right)
-\left(\dot{x^s}\dot{x^s}\right)\frac{\partial}{\partial x^k}
\left(\frac{1}{r}\right)-
\frac{3}{2r^5}\left(x^s\dot{x^s}\right)^2x^k\right].\eqno(85b)$$
Consequently we obtain from the geodesic equation (84) the following equation
of motion for the planet:
$$\ddot{x^k}-\frac{r_s}{2}
\frac{\partial}{\partial x^k}\left(\frac{1}{r}\right)$$
$$=r_s\left[\left(\dot{x^s}\dot{x^s}\right)\frac{\partial}{\partial x^k}
\left(\frac{1}{r}\right)-\frac{r_s}{2r}\frac{\partial}{\partial x^k}
\left(\frac{1}{r}\right)-\dot{x^n}\frac{\partial}{\partial x^n}
\left(\frac{1}{r}\right)\dot{x^k}+
\frac{3}{2r^5}\left(x^s\dot{x^s}\right)^2x^k\right].\eqno(86)$$
Replacing now the derivatives with respect to $x^0$ by those with respect to
$t(\equiv x^0/c)$ in the latter equation, we obtain
$$\ddot{\mbox{\bf x}}-GM\nabla\frac{1}{r}=
r_s\left[\left(\dot{\mbox{\bf x}}^2\right)\nabla
\left(\frac{1}{r}\right)-\frac{GM}{r}\nabla
\left(\frac{1}{r}\right)-\left(\dot{\mbox{\bf x}}\cdot\nabla
\frac{1}{r}\right)\dot{\mbox{\bf x}}+
\frac{3}{2r^5}\left(\mbox{\bf x}\cdot\dot{\mbox{\bf x}}\right)^2\mbox{\bf x}\right],
\eqno(87)$$
where use has been made of the three-dimensional notation.

Hence the equation of motion of the planet differs from the Newtonian one 
since the left-hand side of Eq. (87) is proportional to terms of order of 
magnitude $r_s$ instead of vanishing identically. This correction leads to a
fundamental effect, namely, to a systematically secular change in the
perihelion of the orbit of the planet. 

To integrate the equation of motion (87) we multiply it vectorially by the
radius vector {\bf x}. We obtain
$$\mbox{\bf x}\times\ddot{\mbox{\bf x}}=-r_s\left(\dot{\mbox{\bf x}}\cdot
\nabla\left(1/r\right)\right)
\left(\mbox{\bf x}\times\dot{\mbox{\bf x}}\right).\eqno(88)$$
All other terms in Eq. (87) are proportional to the radius vector {\bf x} and
thus contribute nothing. Equation (88) may be integrated to yield the first 
integral
$$\mbox{\bf x}\times\dot{\mbox{\bf x}}=\mbox{\bf J}e^{-r_s/r}.\eqno(89)$$
Here {\bf J} is a constant vector, the {\it angular momentum} per mass unit of 
the planet. One can easily check that the first integral (89) indeed leads 
back to Eq. (88) by taking the time derivatives of both sides of Eq. (89).

From Eq. (89) we see that the radius vector {\bf x} moves in a plane 
perpendicular to the constant angular momentum vector {\bf J}, thus the
planet moves in a plane similar to the case in Newtonian mechanics. If we now
introduce in this plane coordinates $r$ and $\phi$ to describe the motion of
the planet, the equation of motion (87) consequently decomposes into two
equations.
Introducing now the new variable $u=1/r$, we can then rewrite the equations 
in terms of $u(\phi)$, using
$$\dot{r}=-\frac{u'}{u^2}\dot{\phi},$$
$$\ddot{r}=\frac{2u'^2}{u^3}\dot{\phi}^2-\frac{u''}{u^2}\dot{\phi}^2-
\frac{u'}{u^2}\ddot{\phi},$$
where a prime denotes a differentiation with respect to the angle $\phi$. We
subsequently obtain
$$\ddot{\phi}=2\frac{u'}{u}\dot{\phi}^2-\frac{2GM}{c^2}u'\dot{\phi}^2.$$
A straightforward calculation then gives, using the expression for 
$\ddot{\phi}$,
$$u''+u-GM\left(\frac{u^2}{\dot{\phi}}\right)^2=\frac{GM}{c^2}\left[2u^2-u'^2
-2GMu\left(\frac{u^2}{\dot{\phi}}\right)^2\right]\eqno(90).$$
The latter equation can be further simplified if we use the first integral
$$r^2\dot{\phi}=Je^{-2GM/c^2r}.$$
We obtain
$$\frac{u^2}{\dot{\phi}}=\frac{1}{J}e^{2GMu/c^2},$$
$$\left(\frac{u^2}{\dot{\phi}}\right)^2=\frac{1}{J^2}e^{4GMu/c^2}\approx
\frac{1}{J^2}\left(1+\frac{4GM}{c^2}u\right).$$
Hence, to an accuracy of $1/c^2$, Eq. (90)  gives
$$u''+u-\frac{GM}{J^2}=\frac{GM}{c^2}\left(2u^2-u'^2+2\frac{GM}{J^2}u\right).
\eqno(91)$$

Equation (91) can be used to determine the motion of the planet. The Newtonian 
equation of motion that corresponds to Eq. (91) is one whose left-hand side is
identical to the above equation, but is equal to zero rather than to the terms
on the right-hand side. This fact can easily be seen if one lets $GM/c^2$ go
to zero in Eq. (91). Therefore in the Newtonian limit we have
$$u''+u-\frac{GM}{J^2}\approx 0,\eqno(92)$$
whose solution can be written as
$$u\approx u_0\left(1+\epsilon\cos\phi\right).\eqno(93)$$
Here $u_0$ is a constant, and $\epsilon$ is the eccentricity of the ellipse,
$\epsilon=(1-b^2/a^2)^{1/2},$ where $a$ and $b$ are the semimajor and 
semiminor axes of the ellipse. Using the solution (93) in the Newtonian limit
of the equation of motion (92) then determines the value of the constant 
$u_0$, as $u_0=GM/J^2$. 

To solve the equation of motion (91), we therefore assume
a solution of the form
$$u=u_0\left(1+\epsilon\cos\alpha\phi\right),\eqno(94)$$
where $\alpha$ is some parameter to be determined, and whose value in the 
usual nonrelativistic mechanics is unity. The appearance of the parameter 
$\alpha\neq 1$ in our solution is an indication that the motion of the planet
will no longer be a closed ellipse.

Using the above solution in Eq. (91), and equating coefficients of 
$\cos\alpha\phi$, then gives
$$\alpha^2=1-\frac{2GM}{c^2}\left(2u_0+\frac{GM}{J^2}\right).$$
If we substitute for $GM/J^2$ in the above equation its nonrelativistic value
$u_0$, then the error will be of a higher order. Hence the latter equation can
be written as 
$$\alpha^2=1-\frac{6GM}{c^2}u_0$$
or
$$\alpha=1-\frac{3GM}{c^2}u_0.\eqno(95)$$

Successive perihelia occur at two angles $\phi_1$ and $\phi_2$ when $\alpha
\phi_2-\alpha\phi_1=2\pi$. Since the parameter $\alpha$ is smaller than unity,
we have $\phi_2-\phi_1=2\pi/\alpha>2\pi$. Hence we can write $\phi_2-\phi_1=
2\pi+\Delta\phi$, with $\Delta\phi>0$, or
$$\alpha\left(\phi_2-\phi_1\right)=\alpha\left(2\pi+\Delta\phi\right)=\left(
1-\frac{3GM}{c^2}u_0\right)\left(2\pi+\Delta\phi\right)=2\pi.\eqno(96)$$
As a result there will be an {\it advance} in the perihelion of the orbit of
the planet per revolution given by Eq. (96) or, to first order, by
$$\Delta\phi=6\pi GMu_0/c^2.\eqno(97)$$

The constant $u_0$ can also be expressed in terms of the eccentricity, using
the Newtonian approximation. Denoting the radial distances of the orbit, which
correspond to the angles $\phi_2=0$ and $\phi_1=\pi$, by $r_2$ and $r_1$, 
respectively, we have from Eq. (93),
$$1/r_2=u_0\left(1+\epsilon\right),\mbox{\hspace{5mm}}
1/r_1=u_0\left(1-\epsilon\right).$$ 
Hence since $r_1+r_2=2a$, we obtain (see Fig. 9)
$$2a=r_1+r_2=2/u_0\left(1-\epsilon^2\right),$$
where $a$ is the semimajor axis of the orbit, and therefore
$$u_0=1/a\left(1-\epsilon^2\right).$$
Using this value for $u_0$ in the expression (97) for $\Delta\phi$, we obtain
for the perihelion advance the expression
$$\Delta\phi=\frac{6\pi GM}{c^2a\left(1-\epsilon^2\right)}\eqno(98)$$
in radians per revolution (see Fig. 10). This is the standard general
relativistic formula for the advance of the perihelion.

In the next section we discuss the deflection of a light ray moving in a 
gravitational field.
\section{Deflection of light in a gravitational field}
To discuss the effect of gravitation on the propagation of light signals we
may use the geodesic equation, along with the null condition $ds=0$ at a fixed 
velocity. A light signal propagating in the gravitational field of the Sun, 
for instance, will thus be described by the null geodesics in the cosmological
Schwarzschild field at $dv=0$.   

Using the approximate solution for the cosmological Schwarzschild metric, 
given by Eq. (82a), we obtain
$$g_{\mu\nu}dx^\mu dx^\nu=\left(1-\frac{2GM}{c^2r}\right)c^2dt^2-\left[dx^s
dx^s+\frac{2GM}{c^2}\frac{\left(x^sdx^s\right)^2}{r^3}\right]=0.\eqno(99)$$
Hence we have, to the first approximation in $GM/c^2$, the following 
equation of motion for the propagation of light in a gravitational field:
$$\left(1+\frac{2GM}{c^2r}\right)\left[\left(\dot{x^s}\dot{x^s}\right)+
\frac{2GM}{c^2}\frac{\left(x^s\dot{x^s}\right)^2}{r^3}\right]=c^2,\eqno(100)$$
where a dot denotes differentiation with respect to the time coordinate 
$t(\equiv x^0/c)$.

Just as in the case of planetary motion (see previous section), the motion 
here also takes place in a plane. Hence in this plane we may introduce the 
polar coordinates $r$ and $\phi$. The equation of motion (100) then yields, 
to the first approximation in $GM/c^2$, the following equation in the polar
coordinates:
$$\left(\dot{r}^2+r^2\dot{\phi}^2\right)+\frac{4GM}{c^2}\frac{\dot{r}^2}{r}+
\frac{2GM}{c^2}r\dot{\phi}^2=c^2.\eqno(101)$$
Changing now variables from $r$ to $u(\phi)\equiv 1/r$, we obtain
$$\left[u'^2+u^2+\frac{2GMu}{c^2}\left(2u'^2+u^2\right)\right]
\left(\frac{\dot{\phi}}{u^2}\right)^2=c^2,\eqno(102)$$
where a prime denotes differentiation with respectto  the angle $\phi$.

Moreover we may use the first integral of the motion,
$$r^2\dot{\phi}=Je^{-2GM/c^2r},\eqno(103)$$
in  Eq. (102), thus getting
$$u'^2+u^2+\frac{2GMu}{c^2}\left(2u'^2+u^2\right)=\left(\frac{c}{J}\right)^2
e^{4GMu/c^2}.\eqno(104)$$
Differentiation of this equation with respect to $\phi$ then gives
$$u''+u+\frac{GM}{c^2}\left(2u'^2+4uu''+3u^2\right)=\frac{2GM}{J^2}.
\eqno(105)$$
In Eq. (105) terms have been kept to the first approximation in $GM/c^2$ only.

To solve Eq. (105) we notice that, in the lowest approximation, we have, from
Eq. (104),
$$u'^2\approx\left(\frac{c}{J}\right)^2-u^2,\eqno(106)$$
$$u''\approx-u.\eqno(107)$$
Hence using these approximate expressions in Eq. (105) gives
$$u''+u=\frac{3GM}{c^2}u^2 \eqno(108)$$
for the equation of motion of the orbit of the light ray propagating in a
spherically symmetric gravitational field.

In the lowest approximation, namely, when the gravitational field of the 
central body is completely neglected, the right-hand side of Eq. (108) can be
taken as zero, and therefore $u$ satisfies the equation $u''+u=0$. The 
solution of this equation is a straight line given by 
$$u=\frac{1}{R}\sin\phi,\eqno(109)$$
where $R$ is a constant. This equation for the  straight line shows that $r
\equiv 1/u$ has a minimum value $R$ at the angle $\phi=\pi/2$. If we denote
$y=r\sin\phi$, the straight line (109) can then be described by 
$$y=r\sin\phi=R=\mbox{\rm constant} \eqno(110)$$
(see Fig. 11).

We now use the approximate value for $u$, Eq. (109), in the right-hand side of
Eq. (108), since the error introduced in doing so is of higher order. We 
therefore obtain the following for the equation of motion of the orbit of the
light ray:
$$u''+u=\frac{3GM}{c^2R^2}\sin^2\phi.\eqno(111)$$
The solution of this equation is then given by
$$u=\frac{1}{R}\sin\phi+\frac{GM}{c^2R^2}\left(1+\cos^2\phi\right).\eqno(112)$$
 
Introducing now the Cartesian coordinates $x=r\cos\phi$ and $y=r\sin\phi$, the
above solution can then be written as
$$y=R-\frac{GM}{c^2R}\frac{2x^2+y^2}{\left(x^2+y^2\right)^{1/2}}.\eqno(113)$$
We thus see that for large values of $|x|$ the above solution asymptotically
approaches the following expression:
$$y\approx R-\frac{2GM}{c^2R}|x|.\eqno(114)$$

As seen from Eq. (114), asymptotically,  the orbit of the light ray is 
described by two straight lines in the spacetime. These straight lines make 
angles with respect to the $x$ axis given by $\tan\phi=\pm\left(2GM/c^2R
\right)$ (see Fig. 12).  
The angle of deflection $\Delta\phi$ between the two asymptotes is therefore
given by
$$\Delta\phi=\frac{4GM}{c^2R}.\eqno(115)$$

This is the angle of {\it  deflection} of a light ray in
passing through the gravitational field of a central body, described by the
cosmological Schwarzschild metric. For a light ray just grazing the Sun, 
Eq. (115) gives the value 
$$\Delta\phi=\frac{4GM_{\mbox{\rm Sun}}}{c^2R_{\mbox{\rm Sun}}}=
1.75\mbox{\rm seconds}.$$ 
This is the standard general-relativistic formula. Observations indeed 
confirm this result. One of the latest measurements gives $1.75\pm 0.10$ 
seconds. It is worth mentioning that only general relativity theory and the
present theory predict the correct factor of the deflection of light in the
gravitational field.

In  the next section the gravitational radiation prediction is considered.
\section{Gravitational radiation}
In the following we show that the present theory also predicts gravitational 
radiation, a distinguished result of classical general relativity theory. 
We will not develop a complete theory of gravitational radiation. Rather we
will confine ourselves in showing that the present theory does predict the
phenomenon. This is done in the weak field approximation, as is usually done
in standard general relativity theory.
\subsection{Linear approximation}
For convenience, the coordinate system to be used in the linearized theory 
will be Cartesian, and hence the Minkowskian metric will have the form
$$\eta_{\mu\nu}=\eta^{\mu\nu}=(1,-1,-1,-1,1),\eqno(116)$$
when $c$ and $\tau$ are taken as unity. The gravitational field described by 
the metric tensor $g_{\mu\nu}$ is now called weak if it differs from the
Minkowskian metric tensor by terms which are much smaller than unity,
$$\left|g_{\mu\nu}-\eta_{\mu\nu}\right|\ll 1.\eqno(117)$$
The above condition need not be satisfied in the entire spacetime, and it 
could be valid at a region of it.

We now assume that the metric tensor can be expanded as an infinite series, 
$$g_{\mu\nu}=\eta_{\mu\nu}+\lambda{}_1g_{\mu\nu}+\lambda^2{}_2g_{\mu\nu}+
\cdots,\eqno(118)$$
where $\lambda$ is some small parameter, and we limit ourselves to the 
first-order term ${}_1g_{\mu\nu}$ alone. Hence we can write
$$g_{\mu\nu}\approx \eta_{\mu\nu}+h_{\mu\nu},\eqno(119a)$$
where $h_{\mu\nu}=\lambda{}_1g_{\mu\nu}$. We also expand the contravariant
components of the metric tensor,
$$g^{\mu\nu}\approx \eta^{\mu\nu}+h^{\mu\nu}.\eqno(119b)$$
From the condition $g_{\mu\lambda}g^{\lambda\nu}=\delta_\mu^\nu$ one then is 
able to relate $h^{\mu\nu}$ to $h_{\mu\nu}$ (neglecting nonlinear terms),
$$h^{\mu\nu}=-\eta^{\mu\rho}\eta^{\nu\sigma}h_{\sigma\rho}.\eqno(120)$$
\subsection{The linearized Einstein equations}
We can now derive the linearized Einstein equations. To this end we have to
find the first approximate value of the Einstein tensor, the Ricci tensor, the
Ricci scalar, and the Christoffel symbols. A simple calculation then gives
$$\Gamma^\mu_{\alpha\beta}\approx\frac{1}{2}\eta^{\mu\lambda}\left(
h_{\lambda\alpha,\beta}+h_{\lambda\beta,\alpha}-h_{\alpha\beta,\lambda}\right)
\eqno(121)$$
for the Christoffel symbols and
$$R_{\alpha\beta\gamma\delta}\approx\frac{1}{2}\left(
h_{\alpha\delta,\beta\gamma}+h_{\beta\gamma,\alpha\delta}-
h_{\beta\delta,\alpha\gamma}-h_{\alpha\gamma,\beta\delta}\right)\eqno(122)$$
for the Riemann tensor. Accordingly we have the following expressions for the
Ricci tensor, the Ricci scalar, and the Einstein tensor, respectively:
$$R_{\beta\delta}\approx\frac{1}{2}\eta^{\alpha\gamma}\left(
h_{\alpha\delta,\beta\gamma}+h_{\beta\gamma,\alpha\delta}-
h_{\beta\delta,\alpha\gamma}-h_{\alpha\gamma,\beta\delta}\right)\eqno(123)$$
$$R\approx\eta^{\alpha\gamma}\eta^{\beta\delta}\left(
h_{\alpha\delta,\beta\gamma}-h_{\beta\delta,\alpha\gamma}\right)\eqno(124)$$
$$G_{\mu\nu}\approx-\frac{1}{2}\left[h_{,\mu\nu}+\eta^{\rho\sigma}\left(
h_{\mu\nu,\rho\sigma}-h_{\mu\rho,\nu\sigma}-h_{\nu\rho,\mu\sigma}\right)-
\eta_{\mu\nu}\eta^{\rho\sigma}\left(h_{,\rho\sigma}-\eta^{\alpha\beta}
h_{\rho\sigma,\alpha\beta}\right)\right],\eqno(125)$$
where $h=\eta^{\alpha\beta}h_{\alpha\beta}$.

A simplification in the linearized field equations occurs if we introduce the 
new variables
$$\gamma_{\mu\nu}=h_{\mu\nu}-\frac{1}{2}\eta_{\mu\nu}h,\eqno(126)$$
from which one obtains
$$h_{\mu\nu}=\gamma_{\mu\nu}-\frac{1}{2}\eta_{\mu\nu}\gamma,\eqno(127)$$
with $\gamma=\eta^{\alpha\beta}\gamma_{\alpha\beta}$. 
Introducing the above expressions into the Einstein field equations we obtain 
$$\bigcirc\gamma_{\mu\nu}-\eta^{\alpha\beta}\left(\gamma_{\alpha\mu,\beta\nu}+
\gamma_{\alpha\nu,\beta\mu}\right)+\eta_{\mu\nu}\eta^{\lambda\rho}
\eta^{\alpha\beta}\gamma_{\lambda\alpha,\rho\beta}=-2\kappa T_{\mu\nu}
\eqno(128)$$
for the linearized gravitational field equations. In Eq. (128) the symbol 
$\bigcirc$ is the operator in flat space,
$$\bigcirc f=\eta^{\alpha\beta}f_{,\alpha\beta}=\left(\frac{1}{c^2}
\frac{\partial^2}{\partial t^2}-\nabla^2+\frac{1}{\tau^2}
\frac{\partial^2}{\partial v^2}\right)f.\eqno(129)$$

We can simplify still further the above field equations by choosing 
coordinates in which 
$$\gamma_\mu=\eta^{\rho\sigma}\gamma_{\mu\rho,\sigma}=0.\eqno(130)$$
This is similar to choosing a gauge in solving the wave equation in 
electrodynamics. As a result we finally obtain for the linearized Einstein
equations the following:
$$\bigcirc\gamma_{\mu\nu}=-2\kappa T_{\mu\nu},\eqno(131)$$
along with the supplementary condition
$$\eta^{\rho\sigma}\gamma_{\mu\rho,\sigma}=0,\eqno(132)$$
which solutions $\gamma_{\mu\nu}$ of Eq. (131) should satisfy. Finally we see
from Eq. (131) that a necessary condition for Eq. (132) to be satisfied is 
that
$$\eta^{\alpha\beta}T_{\mu\alpha,\beta}=0,\eqno(133)$$
which is an expression for the conservation of the energy and momentum without 
including gravitation.
\subsection{Gravitational waves}
In vacuum, Eq. (131) reduces to 
$$\bigcirc\gamma_{\mu\nu}=0,\eqno(134)$$
or
$$\left(\nabla^2-\frac{1}{c^2}\frac{\partial^2}{\partial t^2}\right)
\gamma_{\mu\nu}=\frac{1}{\tau^2}\frac{\partial^2\gamma_{\mu\nu}}{\partial v^2}.
\eqno(135)$$
Thus the gravitational field, like the electromagnetic field, propagates in
vacuum with the speed of light. The above analysis also shows the existance of
gravitational waves. 

\newpage
Table 1: The Cosmic Times with respect to the Big Bang, the Cosmic
Temperature and the Cosmic Pressure for each of the Curves in Fig. 1. 
\newline
\begin{tabular}{c r@{}l c r@{.}l r@{}l r@{}l}
\hline\\ 
Curve&\multicolumn{2}{c}{$\Omega_m$}&Time in Units &\multicolumn{2}{c}{Time}&
\multicolumn{2}{c}{Temperature}&\multicolumn{2}{c}{Pressure}\\
No$^\star$.& & & of $\tau$ &\multicolumn{2}{c}{(Gyr)}& &(K)&
\multicolumn{2}{c}{(g/cm$^2$)}\\
\hline\\
\multicolumn{10}{c}{DECELERATING EXPANSION}\\
1&100& &$3.1\times 10^{-6}$&$3$&$87\times 10^{-5}$&1096&&-&4.499\\
2&25& &$9.8\times 10^{-5}$&$1$&$22\times 10^{-3}$&195&.0&-&1.091\\
3&10& &$3.0\times 10^{-4}$&$3$&$75\times 10^{-3}$&111&.5&-&0.409\\
4&5& &$1.2\times 10^{-3}$&$1$&$50\times 10^{-2}$&58&.20&-&0.182\\
5&1&.5&$1.3\times 10^{-2}$&$1$&$62\times 10^{-1}$&16&.43&-&0.023\\
\multicolumn{10}{c}{CONSTANT EXPANSION}\\ 
6&1& &$3.0\times 10^{-2}$&$3$&$75\times 10^{-1}$&11&.15&&0\\
\multicolumn{10}{c}{ACCELERATING EXPANSION}\\ 
7&0&.5&$1.3\times 10^{-1}$&$1$&$62$&5&.538&+&0.023\\
8&0&.245&$1.0$&$12$&$50$&2&.730&+&0.034\\
\hline\\ 
\end{tabular}

$^\star$The calculations are made using Carmeli's cosmological transformation, 
Eq. (2),
that relates physical quantities at different cosmic times when gravity is
extremely weak.

For example, we denote the temperature by $\theta$, and the temperature at the
present time by $\theta_0$, we then have 
$$\theta=\frac{\theta_0}{\sqrt{1-\displaystyle\normalsize\frac{t^2}{\tau^2}}}=
\frac{\theta_0}{\sqrt{1-\displaystyle\normalsize\frac{\left(\tau-T\right)^2}
{\tau^2}}}=\frac{2.73K}{\sqrt{\displaystyle\normalsize\frac{2\tau T-T^2}
{\tau^2}}}=\frac{2.73K}
{\sqrt{\displaystyle\normalsize\frac{T}{\tau}\left(2-\frac{T}{\tau}\right)}},$$
where T is the time with respect to B.B.

The formula for the pressure is given by Eq. (30), 
$p=c(1-\Omega)/8\pi G\tau$.
Using $c=3\times 10^{10}cm/s$, $\tau=3.938\times 10^{17}s$ and $G=6.67\times 
10^{-8}cm^3/gs^2$, we obtain
$$p=4.544\times 10^{-2}\left(1-\Omega\right)g/cm^2.$$
\newpage
\begin{center}{Table 2:
Cosmological parameters in cosmological general  relativity and in
standard theory}\end{center} 

\begin{tabular}{p{35mm}p{35mm}p{35mm}}
\hline\\
&COSMOLOGICAL&STANDARD\\
&RELATIVITY&THEORY\\
\hline\\
Theory type&Spacevelocity&Spacetime\\
Expansion&Tri-phase:&One phase\\
type&decelerating, constant,&\\
&accelerating&\\
Present expansion&Accelerating&One of three\\
&(predicted)&possibilities\\
Pressure&$0.034g/cm^2$&Negative\\
Cosmological constant&$1.934\times 10^{-35}s^{-2}$&Depends\\
&(predicted)&\\
$\Omega_T=\Omega_m+\Omega_\Lambda$&1.009&Depends\\
Constant-expansion&8.5Gyr ago&No prediction\\
occurs at&&\\
Constant-expansion&Fraction of&Not known\\
duration&second&\\
Temperature at&146K&No prediction\\
constant expansion&&\\
\hline
\end{tabular}
\newpage
\begin{center}{FIGURE CAPTIONS}\end{center}
Fig. 1 Hubble's diagram describing the three-phase evolution of the
Universe according to cosmological general relativity theory. Curves (1) to
(5) represent the stages of {\it decelerating} expansion according to $r(v)=
(c\tau/\alpha)\sin\alpha v/c$, where $\alpha^2=(\Omega-1)/2$, $\Omega=\rho/
\rho_c$, with $\rho_c$ a {\it constant}, $\rho_c=3/8\pi G\tau^2$, and $c$ and
$\tau$ are the speed of light and the Hubble time in vacuum (both universal
constants). As the density of matter $\rho$ decreases, the Universe goes over
from the lower curves to the upper ones, but it does not have enough time to
close up to a big crunch. The Universe subsequently goes to curve (6) with
$\Omega=1$, at which time it has a {\it constant} expansion for a fraction of
a second. This then followed by going to the upper curves (7), (8) with 
$\Omega<1$ where the Universe expands with {\it acceleration} according to
$r(v)=(c\tau/\beta)\sinh\beta v/c$, where $\beta^2=(1-\Omega)/2$. Curve no. 8
fits the present situation of the Universe. (Source: S. Behar and M. Carmeli,
Ref. 3)\vspace{3mm}\newline
Fig. 2 Hubble's diagram of the Universe at the present phase of evolution
with accelerating expansion. (Source: S. Behar and M. Carmeli,
Ref. 3)\vspace{3mm}\newline 
Fig. 3 Hubble's diagram describing decelerating, constant and accelerating 
expansions in a logarithmic scale. (Source: S. Behar and M. Carmeli,
Ref. 3)\vspace{3mm}\newline 
Fig. 4 Distance vs. redshift diagram showing the deviation from a constant
toward an accelerating expansion. (Source: A. Riess {\it et al.}, Ref. 12)\vspace{3mm}\newline 
Fig. 5 Relative intensity of light and relative distance vs. redshift.
(Source: A. Riess {\it et al.}, Ref. 12)\vspace{3mm}\newline
Fig. 6 Hubble diagram of SNe Ia minus an empty (i.e., ``empty" $\Omega=0$)
Universe compared to cosmological and astrophysical models. The points are
the redshift-binned data from the HZT (Riess {\it et al.} 1998) and the SCP
(Perlmutter {\it et al.} 1999). The measurements of SN 1997ff are inconsistent
with astrophysical effects which could mimic previous evidence for an 
accelerating Universe from SNe Ia at $z\approx 0.5$. (Source: A. Riess 
{\it et al.}, Ref. 21)\vspace{3mm}\newline
Fig. 7 Same as Fig. 6 with the inclusion of a family of plausible, flat 
$\Omega_\Lambda$ cosmologies. The transition redshift (i.e., the coasting 
point) between the accelerating and decelerating phases  is indicated and is
given as $[2\Omega_\Lambda/\Omega_M]^{1/3}-1$. SN 1997ff is seen to lie 
within the epoch of deceleration. This conclusion is drawn from the result 
that the apparent brightness of SN 1997ff is inconsistent with values of 
$\Omega_\Lambda\geq 0.9$ and hence a transition redshift greater than that of
SN 1997ff. (Source: A. Riess {\it et al.}, Ref. 21) \vspace{3mm}\newline
Fig. 8 Propagation of light in curved spacetime.\vspace{3mm}\newline
Fig. 9 Newtonian limit of planetary motion. The motion is described by a 
closed ellipse if the effect of other planets is completely neglected.
\vspace{3mm}\newline
Fig. 10 Planetary elliptic orbit with perihelion advance. The effect is a 
general relativistic one. The advance of the perihelion is given by $\Delta
\phi$ in radians per revolution, where $\Delta\phi=6\pi GM/c^2a(1-
\epsilon^2)$, with $M$ being the mass of the Sun, $a$ the semimajor axis, and
$\epsilon$ the eccentricity of the orbit of the planet. \vspace{3mm}\newline
Fig. 11 Light ray  when the effect of the central body's gravitational field
is completely neglected. The light ray then moves along the straight line
$y=r\sin\phi=R=\mbox{\rm constant}$, namely, $u=1/r=(1/R)\sin\phi$ 
\vspace{3mm}\newline
Fig. 12 Bending of a light ray in the gravitational field of a spherically 
symmetric body. The angle of deflection $\Delta\phi=4GM/c^2R$, where $M$ is 
the mass of the central body and $R$ is the closest distance of the light 
ray from the center of the body.\vspace{3mm}\newline
\newpage
\begin{figure}[htb]
\centering
\includegraphics[scale=0.8,angle=90]{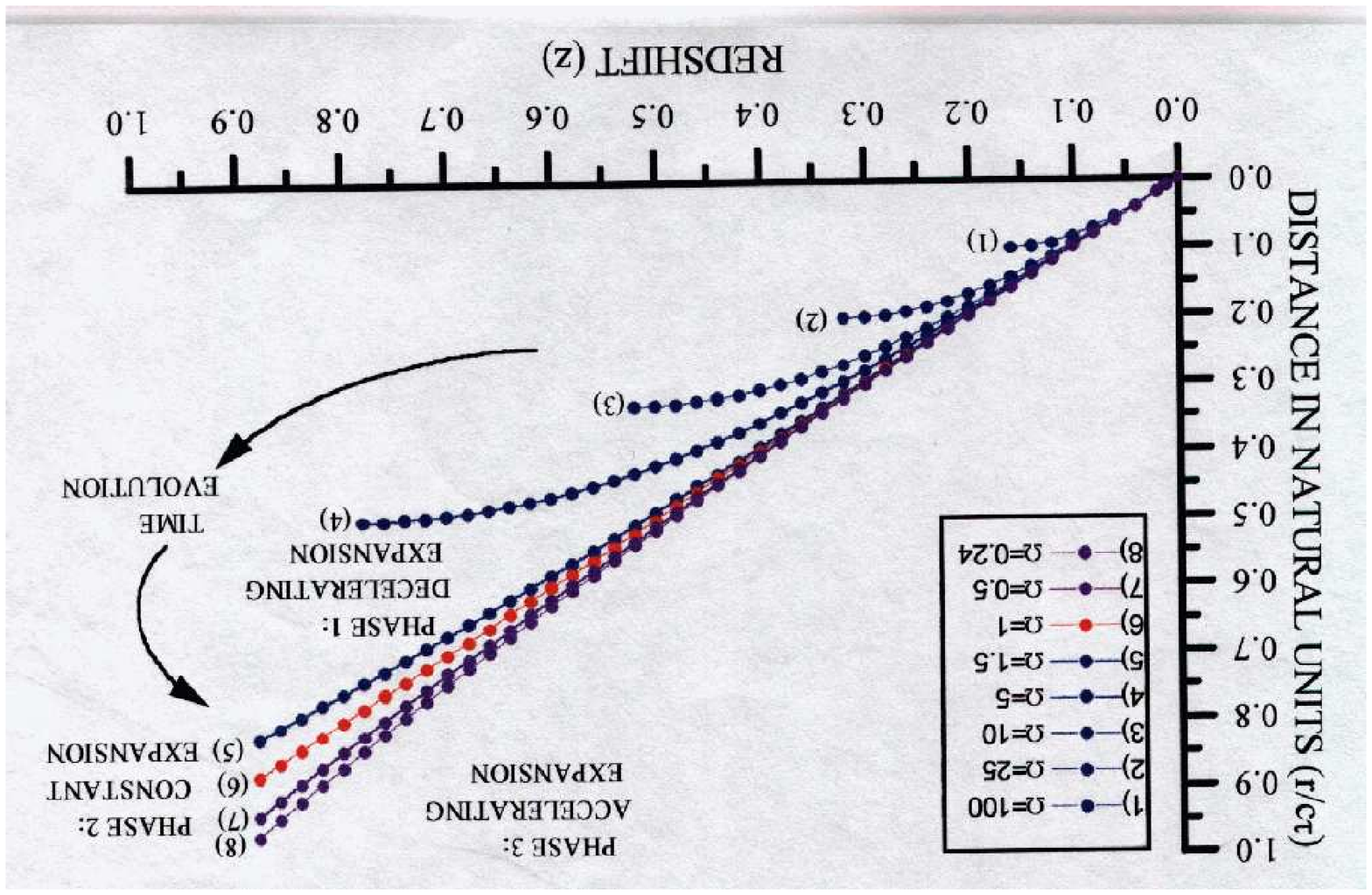}
\end{figure}
\begin{figure}[htb]
\centering
\includegraphics[scale=0.8]{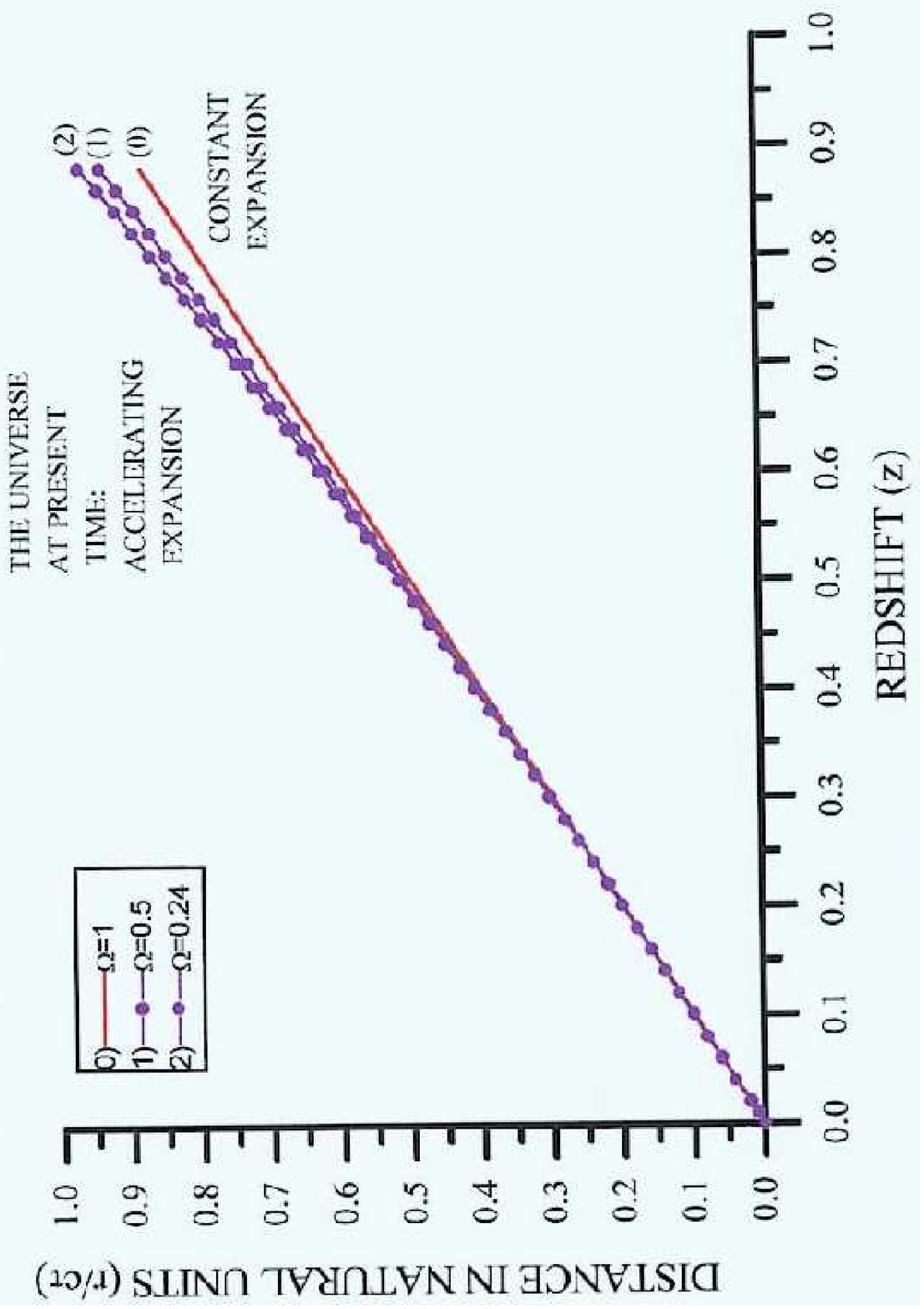}
\end{figure}
\begin{figure}[htb]
\centering
\includegraphics[scale=0.8]{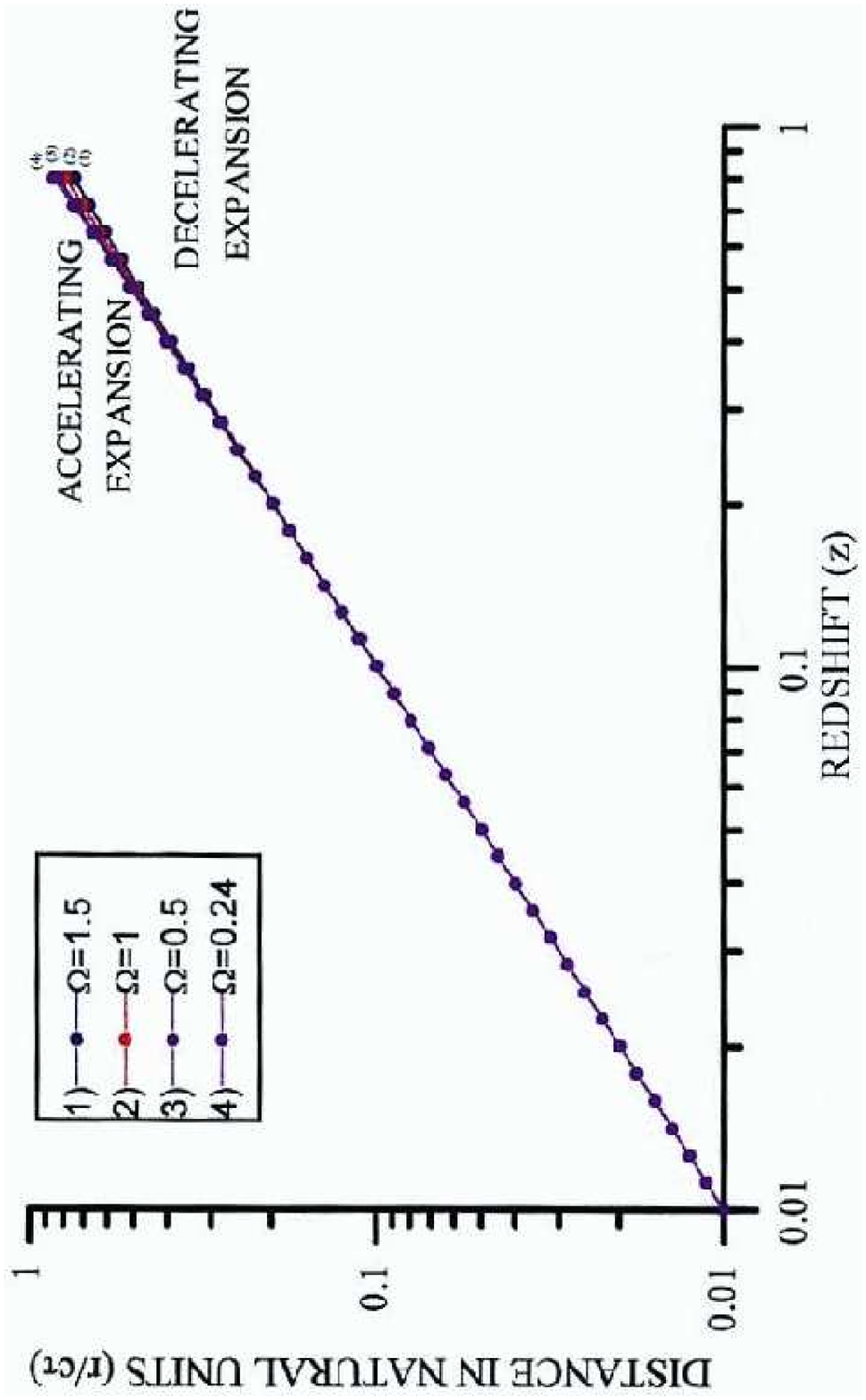}
\end{figure}
\begin{figure}[htb]
\flushleft
\includegraphics[scale=0.8]{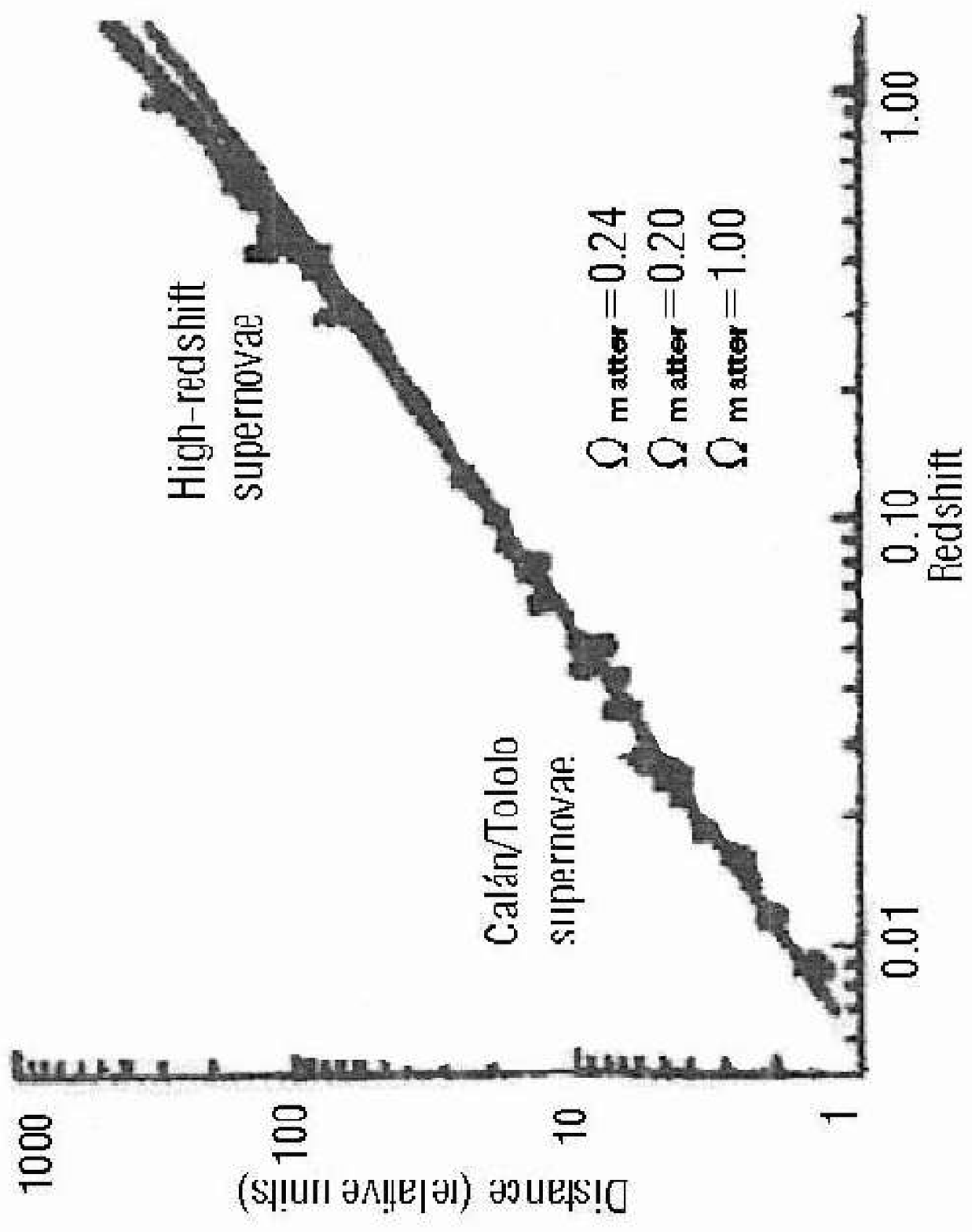}
\end{figure}
\begin{figure}[htb]
\centering
\includegraphics[scale=0.8]{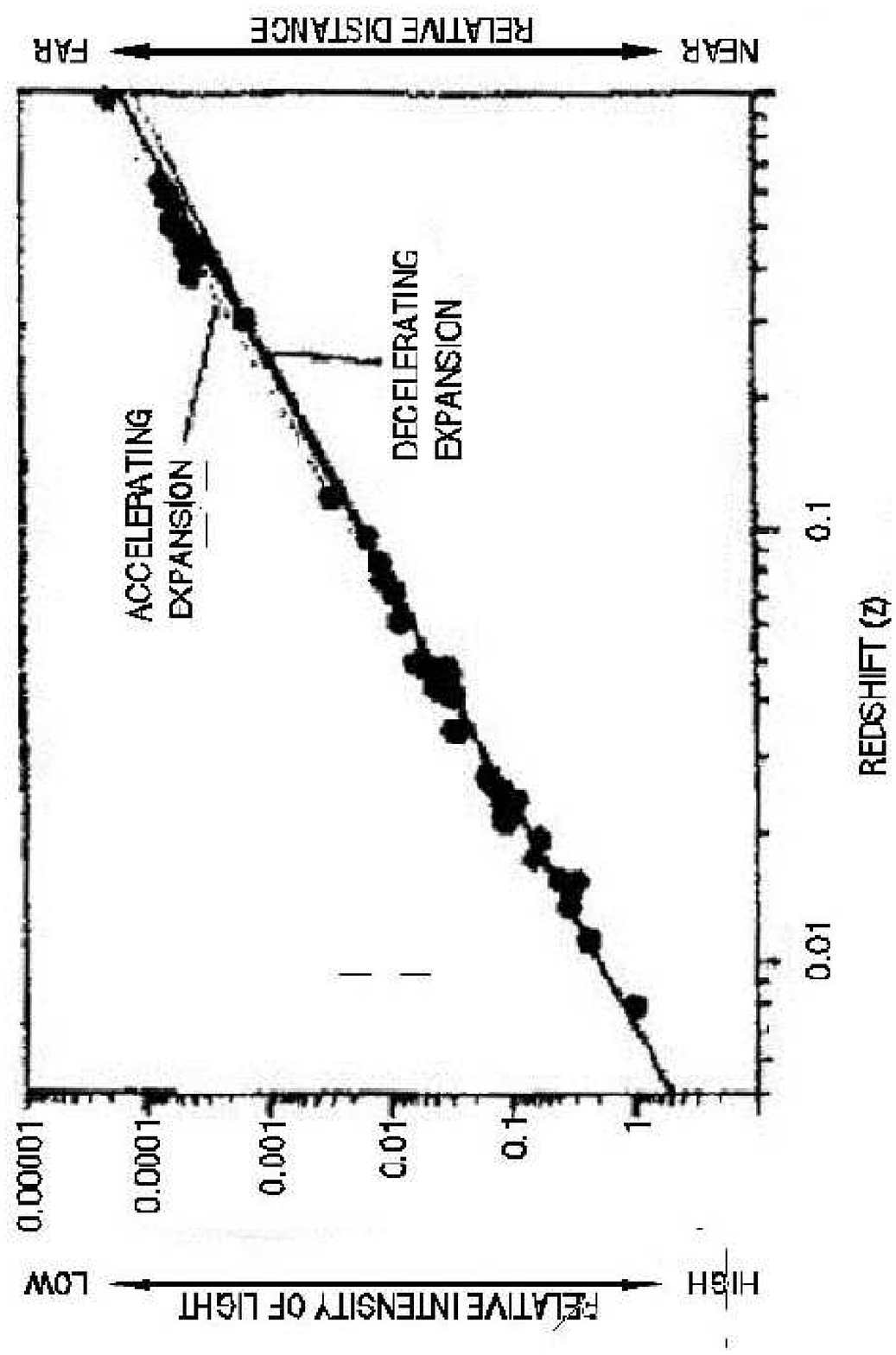}
\end{figure}
\begin{figure}[htb]
\centering
\includegraphics[scale=0.8]{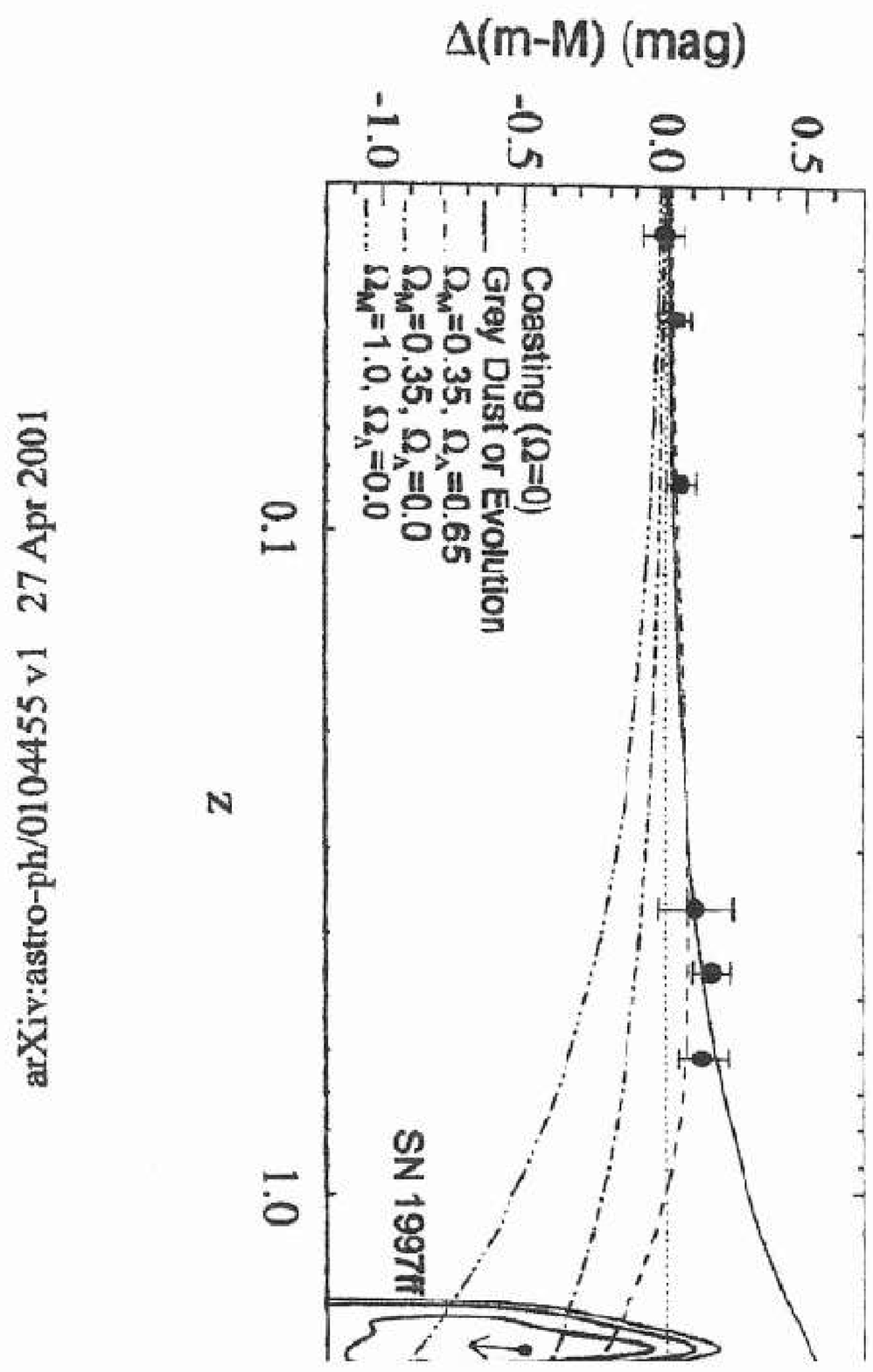}
\end{figure}
\begin{figure}[htb]
\centering
\includegraphics[scale=0.8]{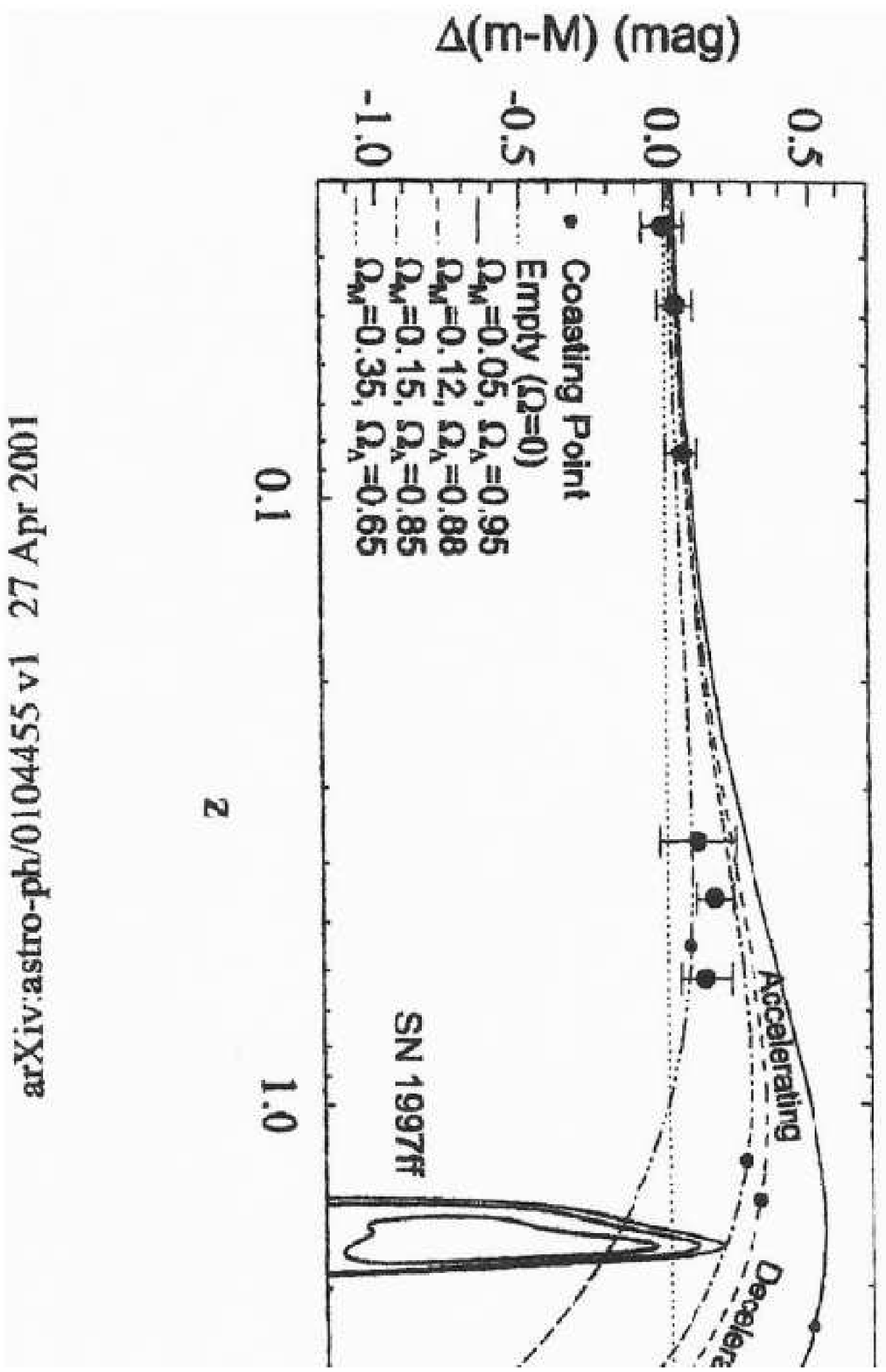}
\end{figure}
\begin{figure}[htb]                                                           |
|\centering                                                                    |
|\includegraphics[scale=0.8]{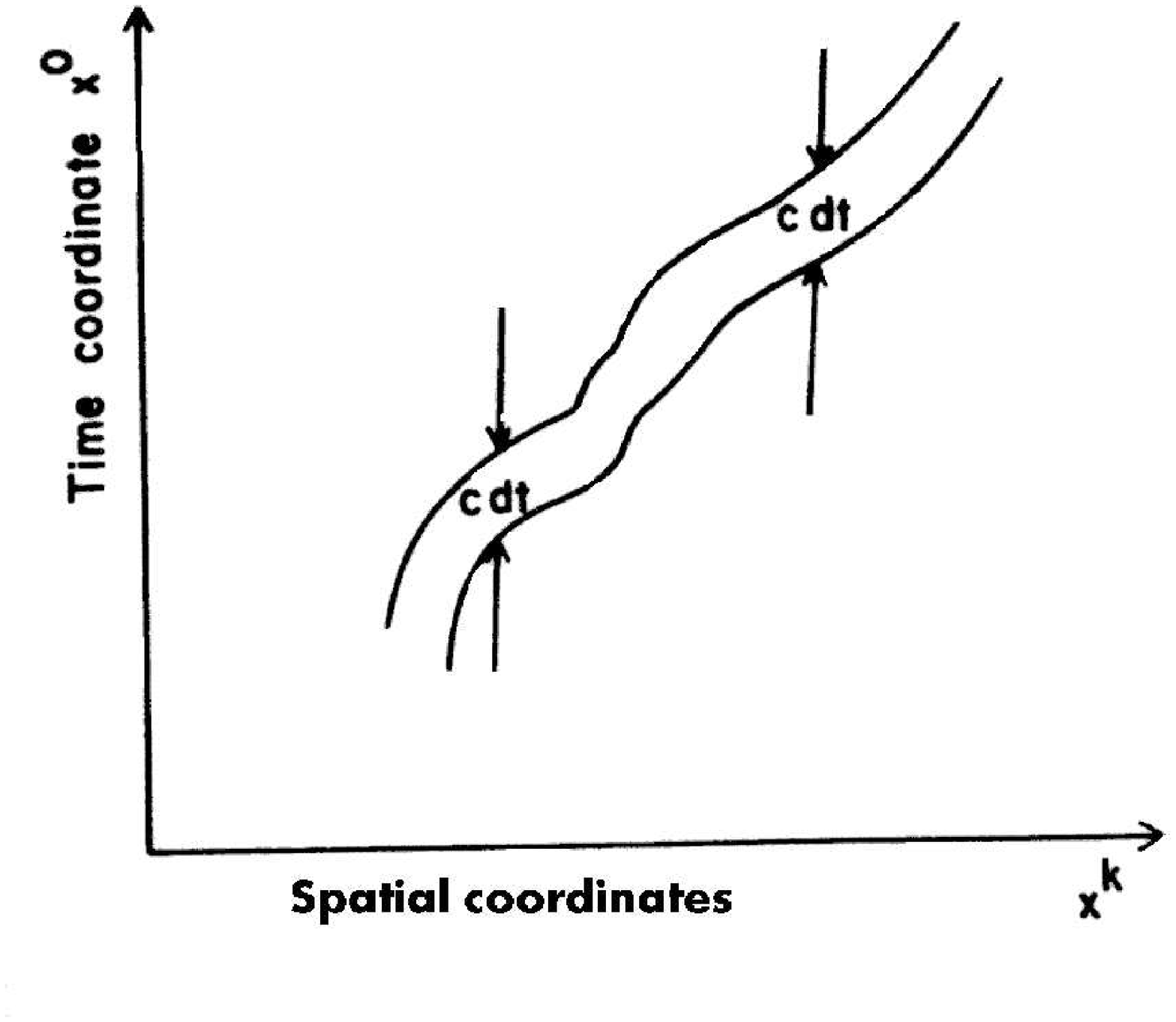}                                       |
|\end{figure} 
\begin{figure}[htb]                                                           |
|\centering                                                                    |
|\includegraphics[scale=0.8]{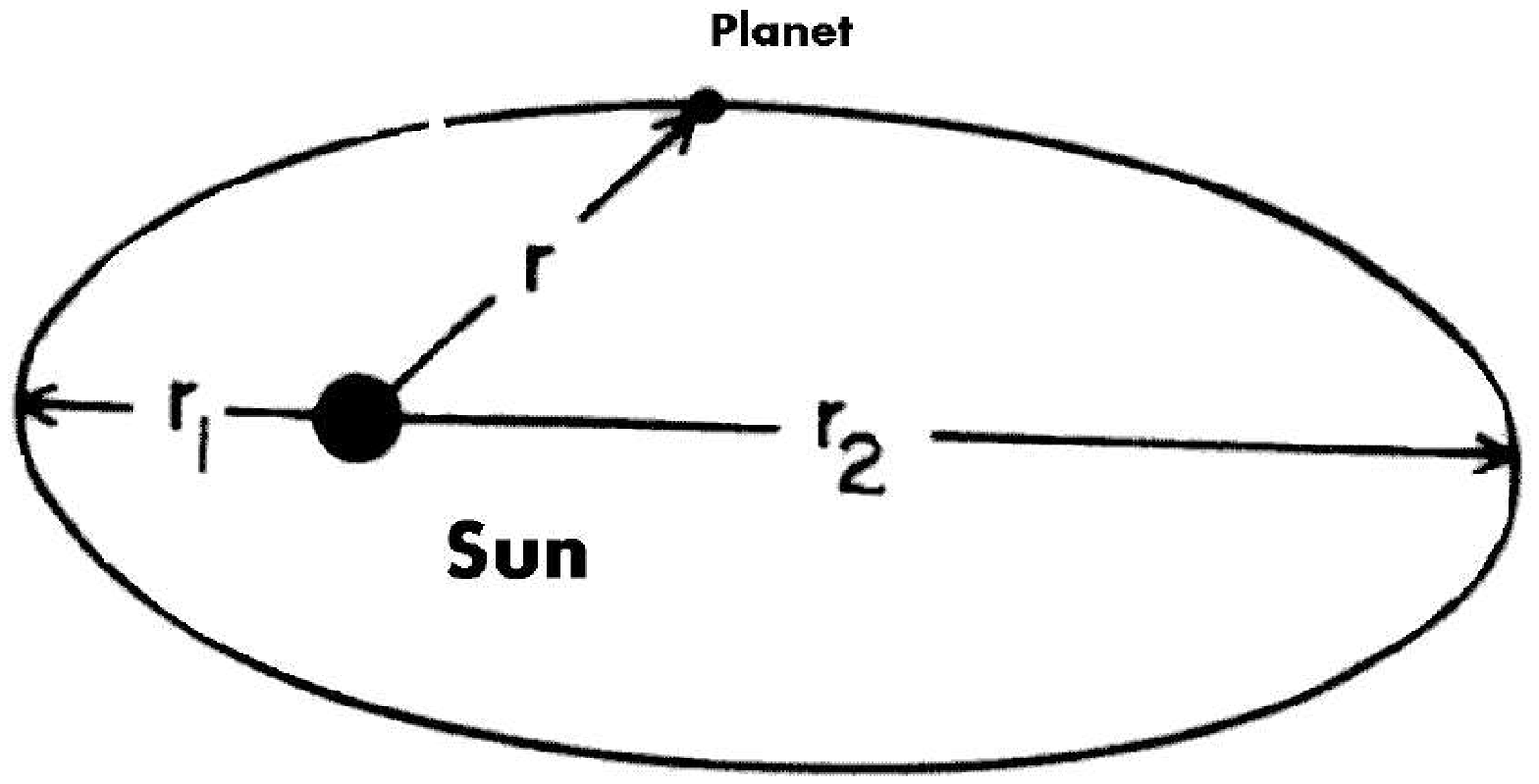}                                       |
|\end{figure} 
\begin{figure}[htb]                                                     |
|\centering
\newpage                                                                    |
|\includegraphics[scale=0.8]{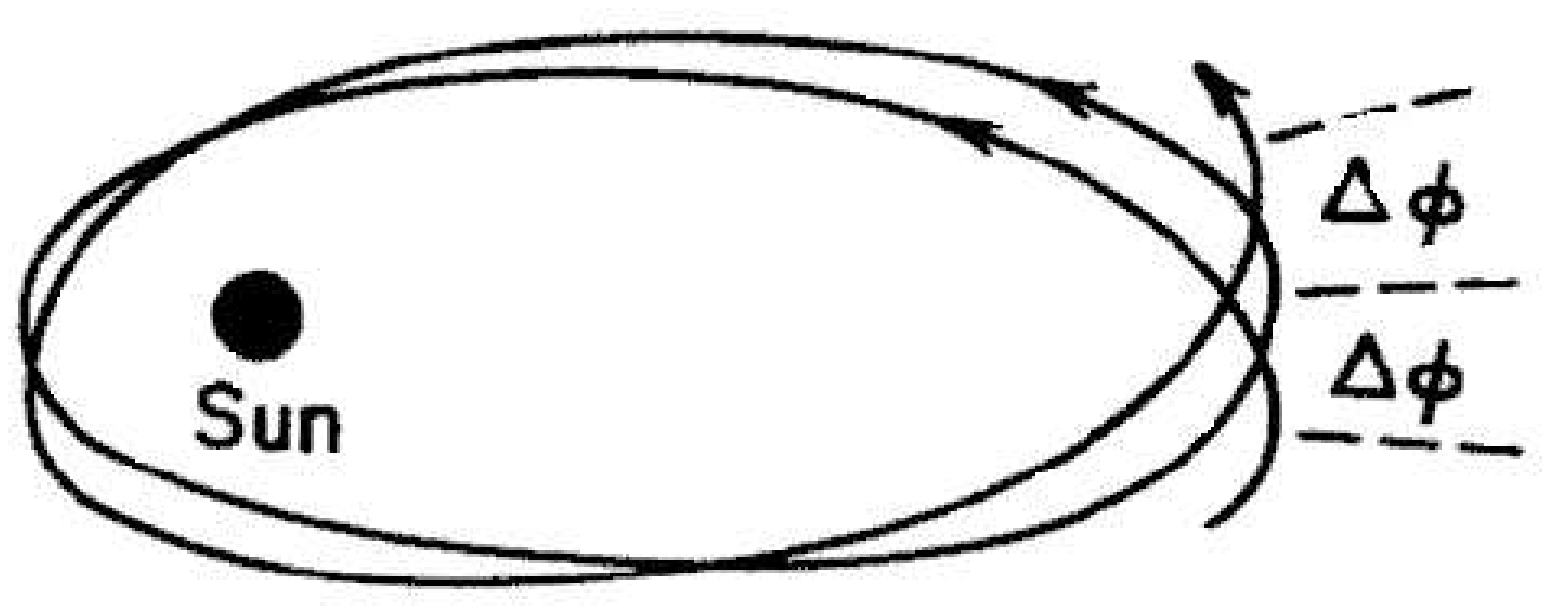}                                       |
|\end{figure} 
\begin{figure}[htb]                                                           |
|\centering                                                                    |
|\includegraphics[scale=0.8]{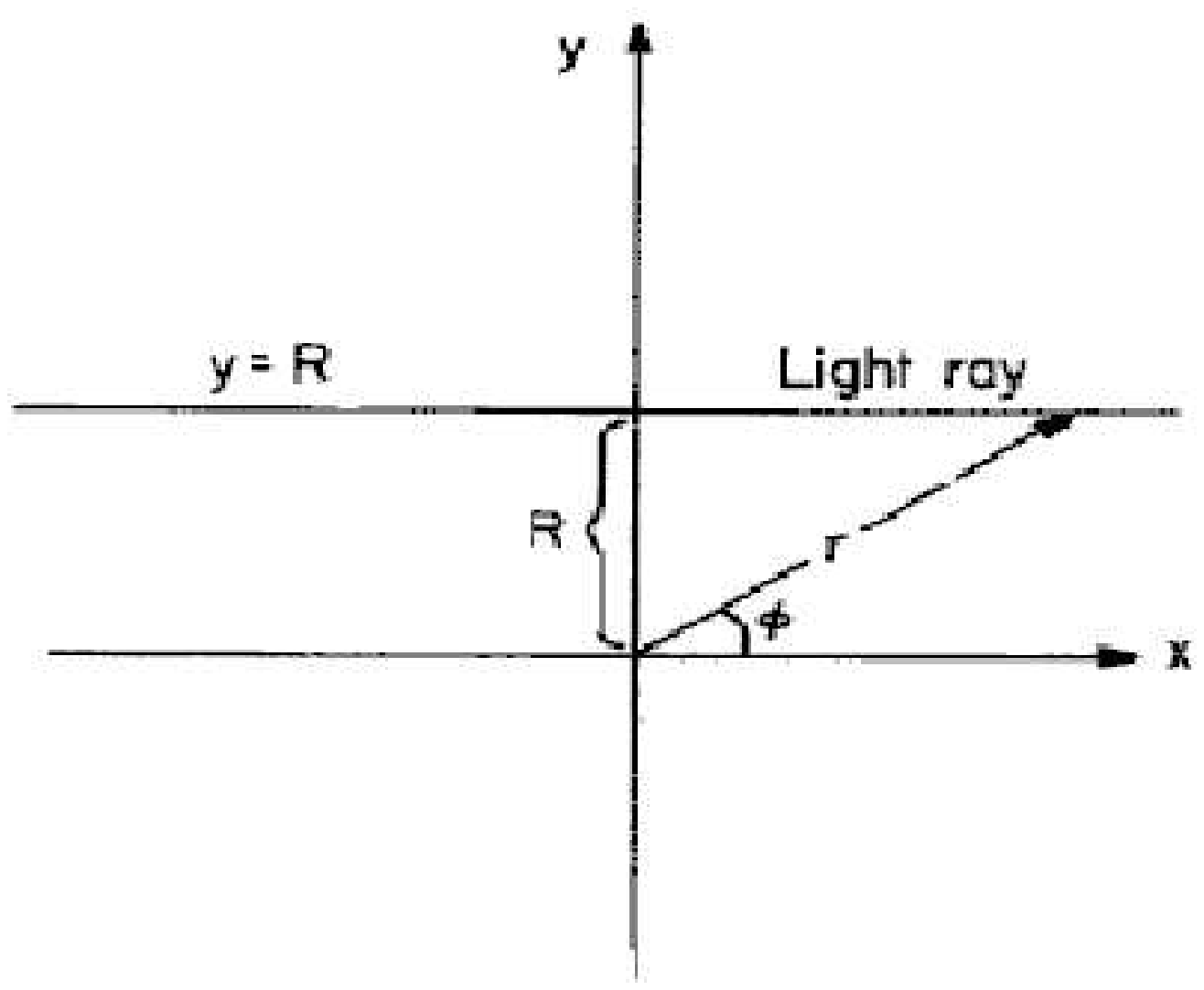}                                       |
|\end{figure} 
\begin{figure}[htb]                                                           |
|\centering                                                                    |
|\includegraphics[scale=0.8]{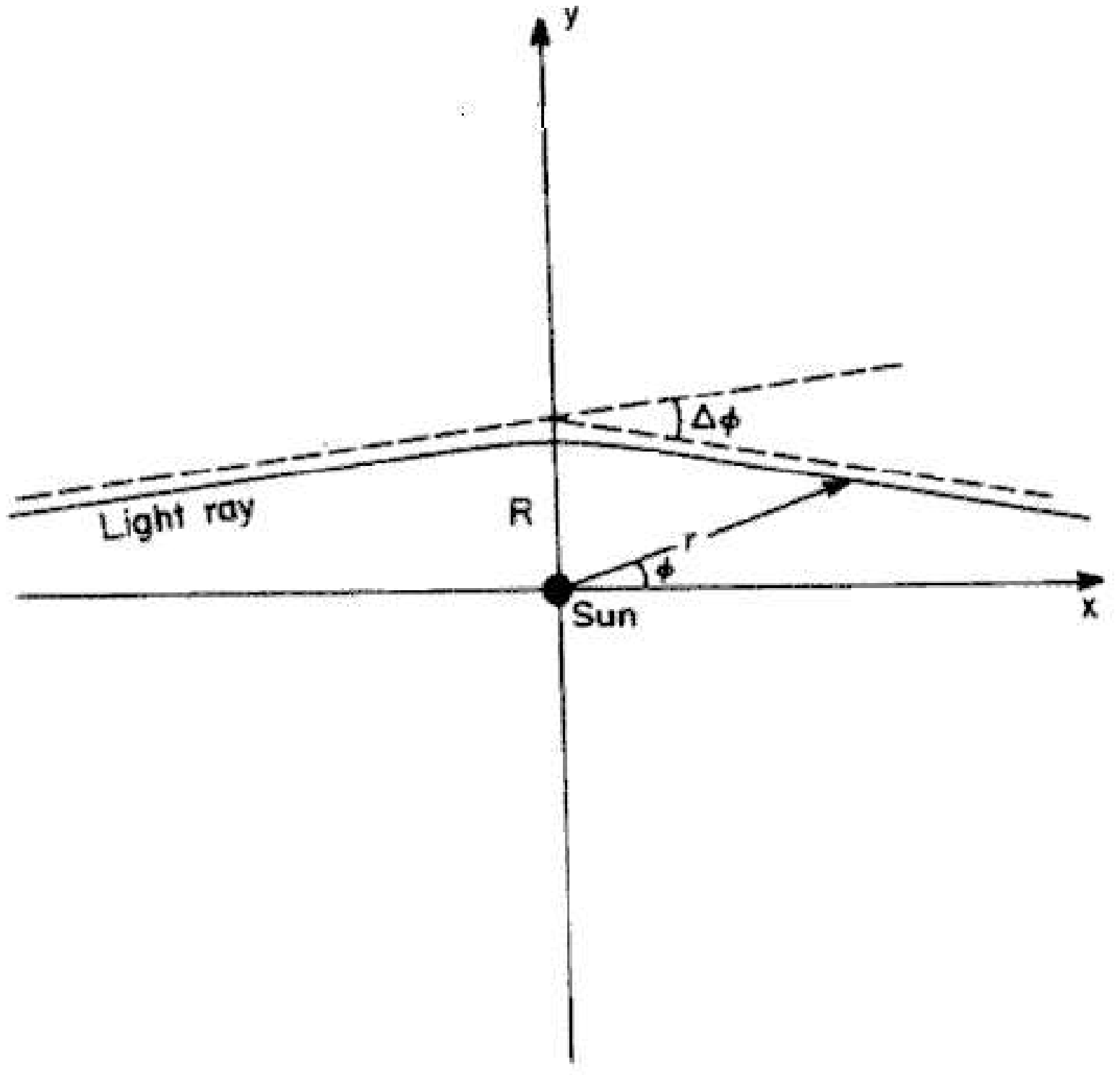}                                       |
|\end{figure} 
\end{document}